\documentclass[letterpaper, 11pt]{article}

\usepackage{fullpage}
\usepackage{amsmath,amsthm}
\usepackage{amssymb, bbm}
\usepackage{epsfig}
\usepackage{stmaryrd}
\usepackage[usenames]{color}
\usepackage[stable]{footmisc}
\usepackage{graphicx}
\usepackage{dsfont}

\def\tr{\operatorname{tr}}
\def\grad{\operatorname{grad}}
\def\div{\operatorname{div}}
\def\Div{\operatorname{Div}}
\def\curl{\operatorname{curl}}
\def\Curl{\operatorname{Curl}}

\numberwithin{equation}{section}

\def\dsp{\def\baselinestretch{1.5}\large\normalsize}
\dsp

\newcommand\mbf[1]{\ensuremath{\mbox{\boldmath$#1$}}}

\theoremstyle{remark}
\newtheorem{rem}{Remark}[section]

\begin{document}

\title{Plastic flow in solids with interfaces}

\author{Anurag Gupta\thanks{Department of Mechanical Engineering, Indian Institute of Technology, Kanpur, UP, India 208016, email: ag@iitk.ac.in (communicating author)}  \and David. J. Steigmann\thanks{Department of Mechanical Engineering, University of California, Berkeley, CA, US 94720, email: steigman@newton.berkeley.edu}}

\maketitle

\begin{abstract}
A non-equilibrium theory of isothermal and diffusionless evolution of incoherent interfaces within a plastically deforming solid is developed. The irreversible dynamics of the interface are driven by its normal motion, incoherency (slip and misorientation), and an intrinsic plastic flow; and purely by plastic deformation in the bulk away from the interface. Using the continuum theory for defect distribution (in bulk and over the interface) we formulate a general kinematical framework, derive relevant balance laws and jump conditions, and prescribe a thermodynamically consistent constitutive/kinetic structure for interface evolution.

\end{abstract}

\noindent {\small \textbf{keywords}: Continuous distribution of dislocations, Finite strain elasto-plasticity, Incoherent interfaces, Interface evolution.}


\section{Introduction}

The motivation for the current work is derived from the processes in material evolution where a moving interface plastically deforms the bulk material region, for example recrystallization and impact induced plasticity \cite{Christian02, HumphreysHatherly04, Meyers94}.  The interface is taken to be a sharp surface separating two distinct regions such as different phases (during phase transition), different crystals (in poly-crystalline materials), and differently oriented single crystals (grain boundary), or a wave front during dynamic deformation. Even after we assume the processes to be isothermal and diffusionless and make simplifying assumptions about the bulk and the interface, the rich dynamics of an interface offers a wide gamut of problems to the material scientist \cite{SuttonBalluffi09} and the mathematician \cite{Gurtin93b} alike. The challenge on one hand is to construct physical models which are amenable to experimental verification and numerical implementation, and on the other hand to analyze the resulting partial differential equations for their well-posedness and properties of the solutions.

The structural nature of the interface is characterized on the basis of its behavior upon relaxation of local stresses. We call an interface incoherent if, upon relaxation, it is locally mapped into two disjoint configurations. Otherwise, we call it coherent. An incoherent interface in an otherwise defect-free solid, after stress relaxation, will result into two separate solids \cite{CermelliGurtin94, LeoSekerka89}. Incoherency is expressed in terms of the incompatibility of the distortion field and leads to interface dislocation density as a smeared-out defect distribution (cf. Bilby and coworkers \cite{Bilby55, Bilbyetal64, BulloughBilby56} and Ch. $2$ of \cite{SuttonBalluffi09}). The interfacial dislocation density along with its bulk counterpart contributes to the Burgers vector for arbitrary closed curves (crossing the interface) in the body. If the interface is coherent then its defect distribution, and consequently its contribution to the net Burgers vector, vanishes identically.

We consider plasticity to be a purely dissipative phenomenon driven by irreversible changes in the microstructure. Even in the bulk, away from the interface, the evolution of plastic flow is a complicated non-linear problem coupled with elasticity and non-local microstructural interactions. Many of the underlying mechanisms remain poorly understood \cite{NabarroDuesbery02} and it is becoming increasingly necessary to develop the theory at a microstructural level. One related concern is to understand the plastic behavior at internal boundaries separating different phases or grains \cite{Kochetal07}. The plastic flow behavior at such boundaries will depend on interface motion, relative distortion of the neighboring grains, and the local shape (for example orientation and curvature). It is clear that a theory for plastic flow at the interface cannot be, in general, modeled along the same lines as the theory associated with the bulk.

Interfaces in solids, with an associated energy density, have been well studied in the context of continuum thermodynamics. We note, in particular, the earlier work done to obtain equilibrium conditions for coherent/incoherent interfaces within elastically deforming solids \cite{JohnsonAlexander86, LarcheCahn78, LarcheCahn85, LeoSekerka89}. These conditions were obtained by minimizing the total energy (bulk and interfacial) under appropriate variations in the domain (see Remark \ref{intcond} for further discussion). Gurtin and coworkers \cite{CermelliGurtin94, Gurtin93, Gurtin00, GurtinStruthers90} extended these results to dynamic interfaces and demonstrated the validity of equilibrium interfacial conditions in wider settings than were previously considered. Their methodology relies upon a version of the virtual work principle where contributions from configurational forces were considered in addition to those from classical forces. All these theories, however, assume the bulk surrounding the interface to be defect-free and thus neglect any possibility of interaction between interfacial and bulk defect densities. They therefore fall short of modeling the behavior of interfaces in a plastically deforming medium. On the other hand, some recent strain gradient plasticity models with interface energies dependent on (infinitesimal) plastic strains \cite{AifantisWillis05, FleckWillis09, Gudmundson04} incorporate interfacial flow rules along the same principles as those in the bulk. These relations furnish boundary data for the bulk equations. While restricting themselves to infinitesimal strains, these models also neglect any coupling with other processes (for example the motion and the relative distortion of the interface).

Our aim is to generalize the above mentioned works by developing a continuum theory for interface evolution in a plastically deforming solid under isothermal and diffusionless conditions. Both the bulk and the interface are assumed to possess a continuous distribution of defects, whose density is related to the local elastic and plastic distortion maps. The role of a relaxed manifold is emphasized in the multiplicative decomposition of deformation gradient in the bulk and at the interface. We restrict our developments to the point of positing specific kinetic laws and therefore stop short of formulating complete boundary-initial-value problems. We however provide a detailed description of the associated kinematics, derive all the necessary balance laws and jump conditions, and use physical and material symmetries to restrict the form of constitutive/kinetic relations. In particular, we derive local dissipation inequalities and highlight the interplay between various dissipative mechanisms and the associated driving forces. The bulk behavior in this paper is modeled after our recent work \cite{Guptaetal07, Guptaetal11} on bulk plasticity.

The central results in this paper are:

$(i)$ The multiplicative decomposition of the interface deformation gradient is equivalently given in terms of two sets of (interfacial) elastic and plastic distortions, cf. \eqref{epd8}. Both of these coincide for coherent interfaces.

$(ii)$ The relation between an incoherency tensor and true interface dislocation densities, cf. \eqref{pwsl38.2} and \eqref{pwsl38.3}.

$(iii)$ The relationship between bulk and interface dislocation densities given in \eqref{pwsl49.9}, which also highlights the fact that interface dislocation density, unlike the bulk dislocation density, does not have a vanishing divergence.

$(iv)$ The dissipation inequality \eqref{tdi17a} arising due to interface motion, plastic flow at the interface, and change in relative distortion across the interface. This inequality demonstrates the underlying coupling between the interface motion, the tangential plastic distortion of the neighboring grains, and the relative tangential distortion of the grains. It provides a starting point for developing kinetic laws governing the out of equilibrium thermodynamic process. Otherwise, in thermodynamic equilibrium, it furnishes additional balance laws to be satisfied at the interface.

$(v)$ The restrictions on the form of kinetic laws, \eqref{inv37.1}-\eqref{inv39}, due to various symmetries in the model.

Our work furnishes the pre-requisite information about kinematics, dissipation, and the basic requirements for constitutive
equations needed for the formulation of complete boundary-initial-value problems in the study of dynamic incoherent interfaces within plastically deforming solids.

We have divided this work into three parts. In the first, we prepare the necessary background for studying the thermodynamics of energetic interfaces within a bulk medium. The second part is concerned with the interface dislocation density as a measure of defect distribution over the interface and its relation with the bulk dislocation density. The final part deals with the energetics and kinetics of incoherent interfaces moving within plastically deforming solids. We make certain constitutive assumptions about the nature of interfacial energies and use them to evaluate the net dissipation at the interface. Motivated by the dissipation inequality, and exploiting various symmetries of the physical space and the material, we formulate restrictions on kinetic laws at the interface.

\section{Preliminaries for the theory}

In the following we prepare the ground work for the next two sections. Our discussion on the kinematics and thermodynamics of surfaces, in Subsections \ref{ss} and \ref{bld}, is largely based upon the work of Gurtin and coworkers \cite{CermelliGurtin94, Gurtin93, Gurtin00, GurtinStruthers90} and {\v S}ilhav{\'y} \cite{Silhavy10a, Silhavy10b, Silhavy11}. Our derivation of the interface dissipation inequality, cf. \eqref{di2} or \eqref{di2.1}, is however different and appears to be new. Similar relations were obtained in \cite{CermelliGurtin94, Gurtin93, GurtinStruthers90} within the framework of configurational mechanics \cite{Gurtin00}.

\subsection{Three-dimensional continuum} The translation space of a real three-dimensional
Euclidean point space $\mathcal{E}$ is denoted by $\mathcal{V}$.  Let $Lin$ be the space of linear transformations from $\mathcal{V}$ to $\mathcal{V}$ (second order tensors). The groups of invertible tensors, orthogonal tensors, and rotations are denoted by $InvLin$, $Orth$, and $Orth^+$, respectively. The spaces of symmetric, symmetric positive definite, and skew tensors are represented by $Sym$, $Sym^+$, and $Skw$, respectively. The determinant and the cofactor of ${\bf A} \in Lin$ are denoted by $J_{A}$ and $\mathbf{A}^{\ast}$, respectively, where $\mathbf{A}^{\ast }=J_{A}\mathbf{A}^{-T}$ if $\mathbf{A}\in InvLin$ (superscripts $T$ and $-1$ denote the transpose and the inverse, respectively, and $\mathbf{A}^{-T} = ({\bf A}^{-1})^T$). The space $Lin$ is equipped
with the Euclidean inner product and norm defined by $\bf{A\cdot B=\,}\tr(
\bf{AB}^{T})$ (${\bf B} \in Lin$)  and $\left\vert \bf{A}\right\vert ^{2}=\bf{A\cdot
A}$, respectively, where $\tr(\cdot)$ is the trace operator.

We use both indicial as well as bold notation to represent vector and tensor fields. The components in the indicial notation are written with respect to the three-dimensional Cartesian coordinate system. Indices are denoted with roman alphabets appearing as subscripts. Summation is assumed for repeated indices unless stated otherwise. Let $e_{ijk}$ be the three-dimensional permutation symbol; it is $1$ if $\{i, j, k\}$ is an even permutation of $\{1,2,3\}$, $-1$ if it is an odd permutation, and $0$ if any index is repeated.

Let $\kappa_r \subset \mathcal{E}$ and $\kappa_t \subset \mathcal{E}$ be the reference configuration and the spatial (or current) configuration with translation spaces $\mathcal{V}_{\kappa_r} \subset \mathcal{V}$ and $\mathcal{V}_{\kappa_t} \subset \mathcal{V}$, respectively. There exists a bijective map ${\mbf \chi}$ between $\kappa_r$ and $\kappa_t$; therefore for every ${\bf X} \in \kappa_r$ and time $t$ we have a unique ${\bf x} \in \kappa_t$ given by
\begin{equation}
{\bf x} = \mbf{\chi} ({\bf X},t).
\end{equation}
We assume $\mbf \chi$ to be continuous but piecewise differentiable over $\kappa_r$ and continuously differentiable with respect to $t$.

The derivative of a scalar valued differentiable function of tensors ${G}:Lin \rightarrow \mathrm{R}$ (where $\mathrm{R}$ is the set of all real numbers) is a tensor $G_{\bf A}$ defined by
\begin{equation}
G({\bf A} + {\bf B}) = G({\bf A}) + G_{\bf A} \cdot {\bf B} + o(\left\vert\bf B\right\vert),
\end{equation}
where  $\frac{o(|\bf B|)}{|\bf B|} \rightarrow 0$ as $|{\bf B}| \rightarrow 0$. Similar definitions can be made for vector and tensor valued differentiable functions (of scalars, vectors, and tensors). In particular, if the domain of a function is $\kappa_r$ we denote the derivative by $\nabla$; and if it is $\kappa_t$ then we write $\grad$ for the derivative. Such functions are called fields. The divergence and the curl of fields, on $\kappa_r$, are defined by (for ${\bf w} \in \mathcal{V}$)
\begin{eqnarray}
&& \Div{\bf w} = \tr (\nabla{\bf w}),~(\Curl{\bf w})\cdot{\bf c} = \Div ({\bf w}\times{\bf c}), \label{der7}
\\
&& (\Div{\bf A})\cdot{\bf c} = \Div({\bf A}^T{\bf c}),~\text{and}~(\Curl{\bf A}){\bf c} = \Curl({\bf A}^T{\bf c}) \label{der10}
\end{eqnarray}
for any fixed ${\bf c}\in \mathcal{V}$. Similar definitions hold for fields on $\kappa_t$; in this case we denote divergence and curl by $\div$ and $\curl$, respectively. The material time derivative is the derivative of a function with respect to time for fixed $\bf X$; we denote it by a superimposed dot.

The particle velocity ${\bf v} \in \mathcal{V}_{\kappa_t}$ is defined as ${\bf v}({\bf X},t) = \dot{\mbf{\chi}}({\bf X},t)$. If $\mbf \chi$ is differentiable at $\bf X$, then the deformation gradient ${\bf F} \in InvLin: \mathcal{V}_{\kappa_r} \rightarrow \mathcal{V}_{\kappa_t}$ exists at $\bf X$ and is given by ${\bf F} = \nabla \mbf{\chi}$. We assume $\bf v$ and $\bf F$ to be piecewise continuously differentiable over $\kappa_r$; they (and their derivatives) are allowed to be discontinuous only across the singular surface.

\subsection{\label{ss} Singular surface}

A singular surface (or interface) is a two dimensional manifold in the interior of $\kappa_r$ (or $\kappa_t$) across which various fields (and their
derivatives) may be discontinuous, which otherwise are continuous in the body. A singular surface in $\kappa_r$ is given by
\begin{equation}
{S}_r = \{{\bf X} \in \kappa_r : \phi({\bf X},t) = 0\}, \label{s1}
\end{equation}
where $\phi$ is a continuously differentiable function. The unit normal to
the surface and the normal velocity are defined by
\begin{eqnarray}
&& {\mathbb N}({\bf X},t) = \frac{\nabla {\phi}}{|\nabla {\phi}|}~ \text{and} \nonumber \\
&& {U}({\bf X},t) = - \frac{\dot{\phi}}{|\nabla {\phi}|}~,
\label{s2}
\end{eqnarray}
respectively; the derivatives being evaluated at the surface. The projection tensor $\mathbbm{1}({\bf X},t)$ which map vectors in $\mathcal{V}_{\kappa_r}$ to vectors in $T_{S_r}({\bf X}) \subset \mathcal{V}_{\kappa_r}$, where $T_{S_r}({\bf X})$ is the tangent space at ${\bf X} \in S_r$ such that $\mathbb{N} \perp T_{S_r}({\bf X})$, is given by
\begin{equation}
\mathbbm{1} = {\bf 1} - {\mathbb N} \otimes {\mathbb N},
\end{equation}
where ${\bf 1}$ is the identity tensor in $Lin$. Note that $\mathbbm{1}^T = \mathbbm{1}$ and $\mathbbm{1}\mathbbm{1} = \mathbbm{1}$.

The jump in a discontinuous field (say $\Psi$) is defined on the singular
surface and is denoted by
\begin{equation}
\llbracket \Psi \rrbracket = \Psi^+ - \Psi^-, \label{s0.1}
\end{equation}
where $\Psi^+$ and $\Psi^-$ are the limit values of $\Psi$ as one
approaches the singular surface from either side. The $+$
side is the one into which the normal to the
surface points. Let $\Phi$ be another piecewise continuous field. The following relation can be verified by direct substitution using \eqref{s0.1}:
\begin{equation}
\llbracket \Phi \Psi \rrbracket = \llbracket \Phi \rrbracket \langle\Psi\rangle + \langle\Phi\rangle \llbracket \Psi \rrbracket, \label{s0.2}
\end{equation}
where
\begin{equation}
\langle\Psi\rangle = \frac{\Psi^+ + \Psi^-}{2}. \label{s0.3}
\end{equation}

\paragraph{Derivatives on the surface} We first introduce the general idea of derivatives on manifolds embedded in a higher dimensional space (see for example \cite{Silhavy10a, Silhavy11}). Let $\mathcal{M} \subset Lin$ be a manifold in the space of tensors. The derivative of a scalar valued differentiable function of tensors ${g}:\mathcal{M} \rightarrow \mathrm{R}$ is a tensor ${g}_{\mathbb A}$ defined by (for $\{{\mathbb A}, {\mathbb B}\} \in \mathcal{M}$)
\begin{equation}
g({\mathbb A} + {\mathbb B}) = g({\mathbb A}) + g_{\mathbb A} \cdot {\mathbb B} + o(|\mathbb B|), ~\text{such that}~  g_{\mathbb A}P({\mathbb A}) =  g_{\mathbb A}, \label{deronsur}
\end{equation}
where $\frac{o(|\mathbb B|)}{|\mathbb B|} \rightarrow 0$ as $|{\mathbb B}| \rightarrow 0$, and $P({\mathbb A})$ is the orthogonal projection onto the tangent space of $\mathcal{M}$ at ${\mathbb A}$. Similar definitions can be made for vector and tensor valued functions on manifolds.

Let $\mathbbm{f}$, $\mathbbm{v}$, and $\mathbb{A}$  denote a scalar, vector, and tensor valued field, respectively, on $S_r$. They are differentiable at ${\bf X} \in S_r$ if they have extensions $f$, $\bf v$, and $\bf A$ to a neighborhood of $\bf
X$ in $\kappa_r$ which are differentiable at $\bf X$.  The surface gradients of $\mathbbm{f}$, $\mathbbm{v}$, and $\mathbb{A}$ at
${\bf X} \in S_r$ are defined by
\begin{eqnarray}
&& \nabla^S \mathbbm{f}({\bf X}) =  \mathbbm{1}({\bf
X}) \nabla f ({\bf X}), \label{sd1.1}
\\
&& \nabla^S \mathbbm{v}({\bf X}) =  \nabla {\bf v} ({\bf X})\mathbbm{1}({\bf
X}),~\text{and} \label{sd1.2}
\\
&& \nabla^S \mathbb{A}({\bf X}) =  \nabla {\bf A} ({\bf X})\mathbbm{1}({\bf
X}). \label{sd1.3}
\end{eqnarray}
In the rest of the paper we will use the same symbol for both the surface field and its extension.
We define the surface
divergence of $\mathbbm{v}$ as a scalar field $\Div^S \mathbbm{v}$; and of $\mathbb{A}$ as a
vector field $\Div^S \mathbb{A}$ given by
\begin{eqnarray}
&& \Div^S \mathbbm{v} = \tr(\nabla^S \mathbbm{v})~\text{and} \nonumber
\\
&& {\bf c} \cdot \Div^S \mathbb{A} = \Div^S (\mathbb{A}^T {\bf c})
\label{sd2}
\end{eqnarray}
for a fixed ${\bf c}\in\mathcal{V}$. Moreover, we call $\mathbbm{v}$ (or $\mathbb{A}$)
tangential if $\mathbbm{1}\mathbbm{v} = \mathbbm{v}$ ($\mathbbm{1}\mathbb{A} = \mathbb{A}$) and
$\mathbb{A}$ superficial if $\mathbb{A}\mathbbm{1} =
\mathbb{A}$.

Define the curvature tensor $\mathbb{L}$ by
\begin{equation}
\mathbb{L} = -\nabla^S {\mathbb N}. \label{sd3}
\end{equation}
It is straightforward to verify that $\mathbb{L} = \mathbb{L}^T$ (use \eqref{s2}$_1$) and $\mathbb{L}{\mathbb N} = {\bf 0}$. Therefore, $\mathbb N$ is an eigenvector of $\mathbb{L}$ with zero eigenvalue. Since $\mathbb{L}$ is symmetric, the spectral theorem implies that it has three real eigenvalues with mutually orthogonal eigenvectors. Let the two nontrivial eigenvalues be $\zeta_1$ and $\zeta_2$ with eigenvectors in $T_{S_r}$. The mean and the Gaussian curvature associated with the surface are defined as
\begin{equation}
H = \frac{1}{2} (\zeta_1 + \zeta_2)~\text{and}~K=\zeta_1 \zeta_2, \label{sd3.1}
\end{equation}
respectively.

A function ${\mbf \varphi}:(t-\varepsilon,t+\varepsilon)\rightarrow
\kappa_r$, $\varepsilon > 0$ is said to be a normal curve
through ${\bf X}\in S_r$ at time $t$ if for each $\tau \in
(t-\varepsilon,t+\varepsilon)$, ${\mbf \varphi}(\tau) \in S_r$ and
\begin{equation}
{\mbf \varphi}'(\tau) = U ({\mbf \varphi}(\tau),\tau) {\mathbb N}({\mbf
\varphi}(\tau),\tau), \label{sd4}
\end{equation}
where the superscript prime denotes the derivative with respect to the scalar argument. Define the normal time derivative of a field on $S_r$ by (cf. $\S \text{II}.3$ of \cite{Thomas61} and $\S 179$ of \cite{TruesdellToupin60})
\begin{equation}
\mathring{\mathbbm v}({\bf X},t) = \frac{d{\mathbbm v}({\mbf
\varphi}(\tau),\tau)}{d\tau}\Big|_{\tau = t}. \label{sd5}
\end{equation}
It represents the rate of change in $\mathbbm{v}$ with respect to an observer sitting on $S_r$ and moving with the normal velocity $U\mathbb{N}$ of the interface. As an example, on differentiating \eqref{s2}$_1$ and using the definitions for surface divergence and normal time derivative, we obtain
\begin{equation}
\mathring{\mathbb N} = -\nabla^S U. \label{sd5.1}
\end{equation}
Therefore, evolving surfaces $S_r$ are parallel if and only if $U$ is constant over $S_r$ at any fixed time.

\paragraph{Compatibility conditions} The continuity of deformation field ${\mbf \chi} ({\bf X},t)$ across $S_r$ furnishes the following jump conditions for the deformation gradient and the velocity field (cf. Ch. II of \cite{Thomas61} and Ch. C of \cite{TruesdellToupin60}):

\begin{eqnarray}
\llbracket {\bf F}\rrbracket = {\bf k} \otimes {\mathbb N} \label{cc5}~ \text{and}
\\
\llbracket {\bf v}\rrbracket  + U\llbracket {\bf F}\rrbracket {\mathbb N}  = {\bf 0} ~  \forall {\bf X} \in S_r, \label{cc13}
\end{eqnarray}
where ${\bf k} \in \mathcal{V}_{\kappa_t}$ is arbitrary. For $U \neq 0$ these relations can be combined to eliminate $\bf k$,
\begin{equation}
 U\llbracket {\bf F}\rrbracket =  - \llbracket {\bf v}\rrbracket  \otimes {\mathbb N}. \label{cc14}
\end{equation}

\paragraph{Singular surface in the current configuration} The image of the singular surface $S_r$ in the current configuration is given by
\begin{equation}
{s}_t = \{{\bf x} \in \kappa_t: \psi({\bf x},t) = 0, \text{with}~ \psi({\mbf \chi} ({\bf X}, t),t) = \phi({\bf X},t) \}. \label{sd14}
\end{equation}
The scalar function $\psi$ is continuous but, in general, only piecewise differentiable with respect to its arguments. The derivatives of $\psi$ can suffer jump discontinuities at $s_t$. Differentiate $\psi({\mbf \chi} ({\bf X}, t),t) = \phi({\bf X},t)$ (away from $s_t$) with respect to $\bf X$ (at fixed $t$) and $t$ (at fixed $\bf X$), and then restrict the result to the surface, to obtain respectively,
\begin{eqnarray}
&& \nabla \phi = ({\bf F}^\pm)^T (\grad \psi)^\pm ~\text{and} \nonumber
\\
&& \dot{\phi} = (\grad \psi)^\pm \cdot {\bf v}^\pm + \left( \frac{\partial \psi}{\partial t} \right)^\pm, \label{sd16}
\end{eqnarray}
where $\pm$ indicates that either of $+$ or $-$ limit of the field can be used to satisfy the equation (due to smoothness of $\phi$ across the singular surface), and $\frac{\partial \psi}{\partial t}$ indicates the partial derivative of $\psi$ with respect to $t$ at fixed ${\bf x}$. Substitute \eqref{s2} into \eqref{sd16} to get
\begin{eqnarray}
&& \frac{(\grad {\psi})^\pm}{|(\grad {\psi})^\pm|} = \frac{({\bf F}^\pm)^{-T} {\mathbb N}}{|({\bf F}^\pm)^{-T} {\mathbb N}|} = \frac{({\bf F}^\pm)^* {\mathbb N}}{|({\bf F}^\pm)^* {\mathbb N}|}~\text{and} \nonumber
\\
&& - \frac{1}{|(\grad {\psi})^\pm|} \left( \frac{\partial \psi}{\partial t} \right)^\pm = \frac{(\grad {\psi})^\pm}{|(\grad {\psi})^\pm|} \cdot {\bf v}^\pm + \frac{U}{|({\bf F}^\pm)^{-T} {\mathbb N}|} . \label{sd16.01}
\end{eqnarray}
The compatibility relations \eqref{cc5} and \eqref{cc13} yield the $+$ and $-$ value of the expressions on the right hand sides above identical, cf. \eqref{sd8} below. This leads us to define the normal to the surface $s_t$ and the spatial normal velocity by, cf. \eqref{s2},
\begin{eqnarray}
&& {\mathbbm n} = \frac{(\grad {\psi})^\pm}{|(\grad {\psi})^\pm|}~ \text{and} \nonumber
\\
&& {u} = - \frac{1}{|(\grad {\psi})^\pm|} \left( \frac{\partial \psi}{\partial t} \right)^\pm,
\label{sd15}
\end{eqnarray}
respectively; we obtain
\begin{eqnarray}
&& {\mathbbm n} = \frac{({\bf F}^\pm)^{-T} {\mathbb N}}{|({\bf F}^\pm)^{-T} {\mathbb N}|} = \frac{({\bf F}^\pm)^* {\mathbb N}}{|({\bf F}^\pm)^* {\mathbb N}|}~\text{and} \nonumber
\\
&& u = {\mathbbm n} \cdot {\bf v}^\pm + \frac{U}{|({\bf F}^\pm)^{-T} {\mathbb N}|} . \label{sd17}
\end{eqnarray}

The projection tensor $\bar{\mathbbm{1}}({\bf X},t)$ which map vectors in $\mathcal{V}_{\kappa_t}$ to vectors in $T_{s_t}({\bf x}) \subset \mathcal{V}_{\kappa_t}$, where $T_{s_t}({\bf x})$ is the tangent space at ${\bf x} = {\mbf \chi} ({\bf X},t)$ such that ${\mathbbm n} \perp T_{s_t}({\bf x})$, is given by
\begin{equation}
\bar{\mathbbm{1}} = {\bf 1} - {\mathbbm n} \otimes {\mathbbm n}. \label{sd17.1}
\end{equation}

\paragraph{Surface deformation gradient and normal velocity}
For a continuous motion across the
surface, i.e. $\llbracket{\mbf \chi}({\bf X},t)\rrbracket={\bf 0}$
for ${\bf X} \in S_r$, we define the
surface deformation gradient ${\mathbb F}$ and the surface normal
velocity $\mathbbm v$ on $S_r$ as \cite{GurtinStruthers90, Silhavy10a}
\begin{equation}
{\mathbb F} = \nabla^S {\mbf \chi}~\text{and}~{\mathbbm v} = \mathring
{\mbf \chi}. \label{sd6}
\end{equation}
It is then easy to check that
\begin{equation}
{\mathbb F} = {\bf F}^\pm \mathbbm{1}~\text{and}~{\mathbbm v} = {\bf v}^\pm + U {\bf F}^\pm {\mathbb N}. \label{sd7}
\end{equation}

Tensor $\mathbb{F}$ satisfies $\det{\mathbb F} = 0$, which can be verified using $\eqref{sd7}_1$ and $\det{\mathbbm 1} = 0$. Moreover, we have from \eqref{sd7}$_1$ and \eqref{sd17}$_1$,
\begin{equation}
\mathbb{FN} = {\bf 0} ~\text{and}~ \mathbb{F}^T{\mathbbm n} = {\bf 0}. \label{sd6.1}
\end{equation}
Therefore, $\mathbb{F}\mathbbm{1} = \mathbb{F}$ and $\bar{\mathbbm{1}}\mathbb{F} = \mathbb{F}$. The cofactor ${\mathbb F}^*$ of ${\mathbb F}$ is defined by ${\mathbb F}^*({\bf a}\times{\bf b}) = {\mathbb F}{\bf a}\times {\mathbb F}{\bf b}$ for arbitrary vectors $\{{\bf a},{\bf b}\}\in \mathcal{V}_{\kappa_r}$. Let $\{{\mathbbm t}_1,{\mathbbm t}_2\}\in T_{S_r}({\bf X})$ be two unit vectors such that $\{{\mathbbm t}_1,{\mathbbm t}_2,{\mathbb N}\}$ forms a positively oriented orthogonal basis at $\bf X$. Then
\begin{eqnarray}
{\mathbb F}^*{\mathbb N} = {\mathbb F}^*({\mathbbm t}_1\times{\mathbbm t}_2) &=& {\mathbb F}{\mathbbm t}_1\times {\mathbb F}{\mathbbm t}_2 \nonumber
\\
&=& {\bf F}^\pm{\mathbbm t}_1\times {\bf F}^\pm{\mathbbm t}_2 \nonumber
\\
&=& ({\bf F}^\pm)^*{\mathbb N}, \label{sd8}
\end{eqnarray}
where in the third equality we have used $\eqref{sd7}_1$. On the other hand, employ \eqref{sd6.1}$_1$ to conclude that ${\mathbb F}^*{\mathbbm t}_a = {\bf 0}$ ($a=1,2$) and hence ${\mathbb F}^*{\mathbbm 1} = {\bf 0}$. Therefore, ${\mathbb F}^*$ remains non-zero because $({\bf F}^\pm)^*{\mathbb N}$ does not vanish. According to \eqref{sd8}, $|{\mathbb F}^*{\mathbb N}|$ is equal to the ratio $j$ of the infinitesimal areas (on the singular surface) in the current and the reference configuration. Use ${\mathbb F}^* = {\mathbb F}^* {\bf 1} = {\mathbb F}^*{\mathbb N} \otimes {\mathbb N} $ to write
\begin{equation}
{\mathbb F}^* = {j} ({\mathbbm n} \otimes \mathbb{N}). \label{sd8.1}
\end{equation}
Hence $|{\mathbb F}^*| = |{\mathbb F}^*{\mathbb N}| = j$.

Following Penrose \cite{Penrose54} we define a unique tensor ${\mathbb F}^{-1}$, the pseudoinverse (or the generalized inverse) of $\mathbb F$ , such that
\begin{equation}
{\mathbb F}^{-1}{\mathbb F} = {\mathbbm 1} ~\text{and}~ {\mathbb F}{\mathbb F}^{-1} = \bar{\mathbbm 1}, \label{sd8.11}
\end{equation}
which also satisfies
\begin{equation}
{\mathbb F}^{-1} = \left({\bf F}^\pm\right)^{-1} \bar{\mathbbm 1}, \label{sd8.12}
\end{equation}
as can be checked by direct substitution.

For ${\bf F}^\pm \in InvLin$ there exist unique tensors ${\bf R}^\pm \in Orth^+$ and ${\bf U}^\pm \in Sym^+$ such that ${\bf F}^\pm = {\bf R}^\pm {\bf U}^\pm$. For a non-invertible tensor $\mathbb{F}$ there exists a unique positive semidefinite tensor $\mathbb{U} \in Sym$ and a (non-unique) orthogonal tensor  $\bar{\bf R} \in Orth$ such that $\mathbb{F} = \bar{\bf R}\mathbb{U}$. These statements follow from the polar decomposition theorem for invertible and non-invertible tensors. Recall \eqref{sd7}$_1$ to write $\mathbb{F} = {\bf R}^\pm {\bf U}^\pm \mathbbm{1}$. Tensor $\mathbb{U}$ thus satisfies $\mathbb{U}^2 = \mathbb{F}^T\mathbb{F} = \mathbbm{1} {{\bf U}^\pm}^2 \mathbbm{1}$. Define $\mathbb{R} = \mathbb{F} \mathbb{U}^{-1}$, where $\mathbb{U}^{-1}$ is the pseudoinverse of $\mathbb{U}$ such that ${\mathbb U}^{-1}{\mathbb U} = {\mathbb U} {\mathbb U}^{-1} = {\mathbbm 1}$. Tensor $\mathbb{R}$ is unique and satisfies
\begin{equation}
\mathbb{R}^T\mathbb{R} = \mathbbm{1}~\text{and}~\mathbb{R}\mathbb{R}^T = \bar{\mathbbm{1}}. \label{sd8.18}
\end{equation}
Moreover, tensor $\bar{\bf R}$ in the polar decomposition for $\mathbb{F}$ is related to $\mathbb{R}$ as $\bar{\bf R}\mathbbm{1} = \mathbb{R}$. The expression
\begin{equation}
\mathbb{F} = \mathbb{R}\mathbb{U} \label{sd8.19}
\end{equation}
provides a decomposition for $\mathbb{F}$ into unique tensors.

The surface gradient of normal velocity can be calculated from \eqref{sd7}$_2$
\begin{equation}
\nabla^S {\mathbbm v} =  \mathring{\bf F}^\pm \mathbbm{1} - {\bf F}^\pm {\mathbb N} \otimes \mathring{\mathbb N} - U {\bf F}^\pm {\mathbb L}, \label{sd8.2}
\end{equation}
where, in addition to the definitions of surface gradient and normal time derivative, we have used \eqref{sd3} and $\nabla {\bf v}^\pm = \dot{\bf F}^\pm$. Employ \eqref{sd7}$_1$ and
\begin{equation}
\mathring{\mathbbm 1} = - {\mathbb N} \otimes \mathring{\mathbb N} - \mathring{\mathbb N} \otimes {\mathbb N} \label{sd8.3}
\end{equation}
to rewrite \eqref{sd8.2} as
\begin{equation}
\nabla^S {\mathbbm v} =  \mathring{\mathbb F} \mathbbm{1} - U {\mathbb F} {\mathbb L}. \label{sd8.4}
\end{equation}
Consequently it is only for a flat interface ($\mathbb{L} = {\bf 0}$) that we have $\nabla^S {\mathbbm v} =  \mathring{\mathbb F} \mathbbm{1}$ (compare with $\nabla {\bf v} = \dot{\bf F}$).

\begin{rem} Let $g$ be a scalar function on the interface given by $g = \hat{g}({\mathbb F}, {\mathbb N})$. The arguments of $\hat{g}$ satisfy ${\mathbb F}{\mathbb N} = {\bf 0}$ and $|{\mathbb N}|=1$ and therefore form a submanifold, say $\mathrm{G}$, of $Lin \times \mathcal{V}$. The partial derivatives $\tilde{g}_{\mathbb F}$ and $\tilde{g}_{\mathbb N}$ (with respect to $\mathbb F$ and $\mathbb N$, respectively) are evaluated using an extension $\tilde{g}$ of $\hat{g}$ and restricting the result to $\mathrm{G}$. Extension of $\hat{g}$ is any smooth function defined over $Lin \times \mathcal{V}$ such that it is equal to $\hat{g}$ on $\mathrm{G}$. These partial derivatives lie in the tangent space of $\mathrm{G}$ and hence satisfy (cf. \eqref{deronsur}; for a proof see Appendix B of \cite{Silhavy11})
\begin{equation}
\tilde{g}_{\mathbb F} \mathbb{N} + \mathbb{F} \tilde{g}_{\mathbb N} = {\bf 0}~\text{and}~\mathbb{N} \cdot  \tilde{g}_{\mathbb N} = { 0}. \label{deronsur2}
\end{equation}
In the rest of the paper we will use same notation for the function and its extension.

\end{rem}

\begin{rem}(Derivative of $j$) Use \eqref{sd8} and \eqref{sd8.1} to obtain
\begin{equation}
j^2 = (\mathbb{F}\mathbbm{t}_1 \times \mathbb{F}\mathbbm{t}_2)\cdot(\mathbb{F}\mathbbm{t}_1 \times \mathbb{F}\mathbbm{t}_2), \label{deronsur3}
\end{equation}
where $\mathbbm{t}_1$ and $\mathbbm{t}_2$ are functions of only $\mathbb{N}$; i.e., they are arbitrary orthonormal vectors orthogonal to $\mathbb{N}$. To find partial derivative $j_{\mathbb N}$ fix ${\mathbb F}$ in \eqref{deronsur3} and differentiate it on a one-parameter curve in the space of all unit vectors satisfying ${\mathbb F}{\mathbb N} = {\bf 0}$. Apply the definition of cofactor and use \eqref{sd6.1}$_1$ to get $j_{\mathbb N} = {\bf 0}$. Therefore, by \eqref{deronsur2}, $j_{\mathbb F} \mathbb{N} = {\bf 0}$. On the other hand, differentiating  $j = \left| ({\bf F}^\pm)^*{\mathbb N} \right|$ for fix $\mathbb{N}$ yields
\begin{equation}
j_{\mathbb F}=j \mathbb{F}^{-T}. \label{deronsur4}
\end{equation}
Hence the normal time derivative of $j$ is given by (compare with $\dot{J}_F = J_F \dot{\bf F}{\bf F}^{-1} \cdot {\bf 1}$)
\begin{equation}
 {\mathring j} =  j \mathring{\mathbb F}{\mathbb F}^{-1} \cdot \bar{\mathbbm 1}.  \label{deronsur5}
\end{equation}
\end{rem}

\subsection{\label{bld} Balance laws and dissipation inequality}

Assuming a purely mechanical environment and isothermal heat flow we obtain balance laws for mass and momentum, and the dissipation inequalities both for material points on the interface and away from it. We do not state the balance of energy since it is used, under isothermal conditions, only to calculate the net heat flux during the dissipative process.

\paragraph{Surface divergence theorem and surface transport theorem} In addition to divergence and transport theorems for piecewise smooth fields on $\kappa_r$ (see for example Ch. $3$ of \cite{Silhavy97}) we will repeatedly use the following theorems for fields defined on $S_r$. For a vector field ${\mathbbm w} \in \mathcal{V}_{\kappa_r}$ continuously differentiable on $S \subset S_r$
\begin{equation}
\int_{\partial S} {\mathbbm w}\cdot{\mbf \nu} dL = \int_{S} (\Div^S {\mathbbm w}  + 2H
{\mathbbm w}\cdot{\mathbb N}) dA, \label{st6}
\end{equation}
where ${\mbf \nu} \in T_{S_r}$ is the outer unit normal to the closed curve ${\partial S}$ bounding $S$ such that $\{{\mathbb N}, {\mbf \nu}, {\mathbbm t}\}$ forms a positively oriented orthonormal basis on ${\partial S}$ with $\mathbbm t$ as the tangent vector along ${\partial S}$. Moreover if $\mathbbm{w}$ is tangential, i.e. $\mathbbm{1} \mathbbm{w} = \mathbbm{w}$, then $\mathbbm{w}\cdot{\mathbb N} = 0$ and \eqref{st6} reduces to
\begin{equation}
\int_{\partial  S} {\mathbbm w}\cdot{\mbf \nu} dL = \int_{S} \Div^S {\mathbbm w} dA. \label{st6.1}
\end{equation}

\begin{figure}[t!]
\centering
\includegraphics[width=2.5in]{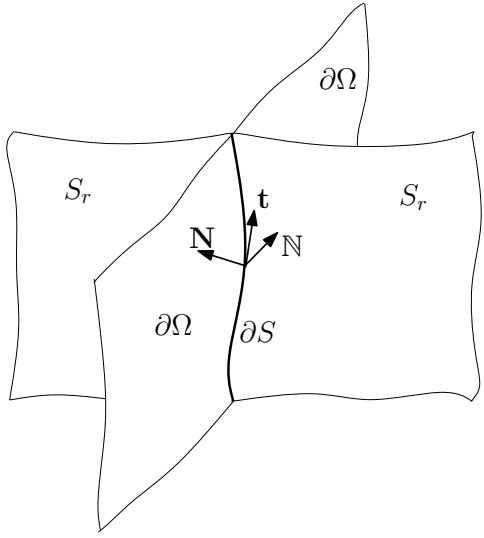}
\caption{A portion of $\partial \Omega$ intersecting with a portion of $S_r$ along $\partial S$. ${\bf N}$ is the outward unit normal to $\partial \Omega$ and ${\mathbb N}$ to $S_r$. ${\bf t}$ is the unit tangent along $\partial S$. Since $\Omega$ is arbitrary we can choose $\partial \Omega$ such that ${\bf t}$ is parallel to ${\mathbbm c}$ and ${\bf N}$ coincides with ${\mbf \nu}$, where ${\mathbbm c} \in T_{S_r}$ is defined in Remark \ref{intcond} and ${\mbf \nu} \in T_{S_r}$ is the outward normal to $\partial S$. This ensures that ${\bf N}\cdot{\mathbb N} = 0$ and ${\mathbbm c}\cdot {\mbf \nu} = 0$ (cf. paragraphs following \eqref{tr5} and \eqref{di2.001}).} \label{intersect}
\end{figure}

The surface transport theorem for an evolving surface $S$ within a fixed region $\Omega$ such that $\partial S \subset \partial \Omega$ is given by \cite{Gurtinetal89}
\begin{equation}
\frac{d}{dt} \int_{S} \mathbbm{w} dA = \int_{S} (\mathring{\mathbbm{w}} - 2UH\mathbbm{w})  dA - \int_{\partial S} \mathbbm{w} U \cot \theta dL, \label{tr5}
\end{equation}
where $\theta = \arccos ({\bf N} \cdot {\mathbb N})$ and ${\bf N}$ is the outward unit normal on $\partial \Omega$. If $\Omega$ is arbitrary then we can always choose $\Omega$ with $\partial \Omega$ such that ${\bf N} \cdot {\mathbb N} = 0$ at all ${\bf X} \in \partial S$ i.e., orient $\partial \Omega$ in such a way that it is orthogonal to $S$ at all points on $\partial S$ (cf. Figure \ref{intersect}). With this choice \eqref{tr5} reduces to
\begin{equation}
\frac{d}{dt} \int_{S} \mathbbm{w} dA = \int_{S} (\mathring{\mathbbm{w}} - 2UH\mathbbm{w})  dA. \label{tr5.1}
\end{equation}
Similar theorems hold for scalar and tensor fields on $S_r$.

\paragraph{Conservation of mass} Assume no net mass transfer in an arbitrary volume of $\kappa_r$. Also assume that there is no additional mass density associated with $S_r$. The statement of conservation of mass then reduces to \cite{Silhavy97}
\begin{equation}
\dot{\rho}_r = 0 ~ \forall {\bf X} \in \kappa_r \setminus S_r,  \label{bl4.3}
\end{equation}
where $\rho_r$ is the referential mass density of the bulk, and
\begin{equation}
U\llbracket\rho_r\rrbracket = 0 ~ \forall {\bf X} \in S_r \label{bl4.4}
\end{equation}
i.e., either the normal velocity vanishes or the referential mass density is continuous across $S_r$.

\begin{figure}[t!]
\centering
\includegraphics[width=2.5in]{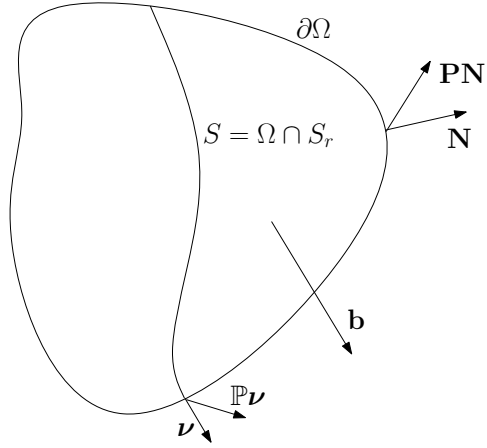}
\caption{Various forces acting on an arbitrary volume $\Omega \subset \kappa_r$.} \label{forces}
\end{figure}

\paragraph{Balance of momentum} The balance laws for linear and angular momentum can be either stated as Euler's postulates or can be deduced from the first law of thermodynamics \cite{Silhavy97}. Let $\Omega$ be a three-dimensional open subset of $\kappa_r$ with boundary $\partial \Omega$ such that $S = \Omega \cap S_r$ is nonempty and $\partial S \subset \partial \Omega$. Let ${\bf N} \in \mathcal{V}_{\kappa_r}$ and ${\mbf \nu} \in T_{S_r}$ be unit vectors normal to $\partial \Omega$ and $\partial S$, respectively. Let ${\bf P} \in Lin$ be the bulk Piola stress and ${\bf b} \in \mathcal{V}$ the specific body force vector. We assume the existence of a contact force between two subsets of $S_r$ along the curve of contact, which can be expressed in terms of a linear map (given by interface Piola stress $\mathbb{P} \in Lin$) acting on the normal to the contact curve \cite{GurtinMurdoch75}. If there are no body forces associated with the singular surface then the balance of linear momentum for $\Omega$ is given by (see Figure \ref{forces} where all the forces are shown)
\begin{equation}
\frac{d}{dt} \int_{\Omega} \rho_r {\bf v} dV = \int_{\partial \Omega} {\bf PN} dA + \int_{\Omega} \rho_r {\bf b} dV + \int_{\partial S} \mathbb{P}{\mbf \nu} dL. \label{bl16}
\end{equation}
Let $\mathbb{P}$ be superficial, i.e. $\mathbb{P}\mathbbm{1} = \mathbb{P}$. This is motivated from the last term of the above equation where $\mathbb{P}\mathbb{N} \otimes \mathbb{N}$ does not contribute to the net force (since $\mathbb{N}$ is orthogonal to ${\mbf \nu}$) and therefore can be assumed to vanish without loss of generality. Integral equation \eqref{bl16} can be localized, using the transport and divergence theorems, to  \cite{Gurtin93, GurtinStruthers90}
\begin{eqnarray}
&& \rho_r \dot{\bf v} = \Div{\bf P} + \rho_r {\bf b} ~ \forall {\bf X} \in \kappa_r \setminus S_r ~\text{and} \label{bl18}
\\
&& U\rho_r \llbracket{\bf v}\rrbracket + \llbracket{\bf P}\rrbracket{\mathbb N} +\Div^S \mathbb{P}  = {\bf 0}  ~ \forall {\bf X} \in S_r, \label{bl19}
\end{eqnarray}
where we have also used \eqref{bl4.3} and \eqref{bl4.4}.

The balance of angular momentum is given by
\begin{equation}
\frac{d}{dt} \int_{\Omega} \rho_r {\bf r} \times
{\bf v} dV = \int_{\partial \Omega} {\bf r} \times
{\bf PN} dA + \int_{\Omega} \rho_r {\bf r} \times
{\bf b} dV + \int_{\partial S} {\bf r} \times
\mathbb{P} {\mbf \nu} dL, \label{bl24}
\end{equation}
where ${\bf r} = {\bf x} - {\bf x}_0$ and ${\bf x}_0 \in \mathcal{E}$ is arbitrary. On using transport and divergence theorems and the equations of balance of mass and linear momentum, it localizes to \cite{Gurtin93, GurtinStruthers90}
\begin{eqnarray}
&& {\bf P}{\bf F}^T = {\bf F}{\bf P}^T ~  \forall {\bf X} \in \kappa_r \setminus S_r ~ \text{and} \label{bl26}
\\
&& {\mathbb P}{\mathbb F}^T = {\mathbb F}{\mathbb P}^T ~  \forall {\bf X} \in S_r. \label{bl26.1}
\end{eqnarray}
Equation \eqref{bl26.1} implies that $\bar{\mathbbm{1}}\mathbb{P} = \mathbb{P}$. Indeed, use $\bar{\mathbbm{1}}\mathbb{F} = \mathbb{F}$ to get $\bar{\mathbbm{1}}{\mathbb P}{\mathbb F}^T = {\mathbb P}{\mathbb F}^T$. The desired result follows upon using \eqref{sd7}$_1$, the invertibility of ${\bf F}^\pm$, and $\mathbb{P}\mathbbm{1} = \mathbb{P}$.

The interface Cauchy stress ${\mathbb T} \in Lin$ is a superficial tensor ($\mathbb{T}\bar{\mathbbm{1}} = \mathbb{T}$) which satisfies
\begin{equation}
\int_{l_t} {\mathbb T}\bar{\mbf \nu} dl = \int_{L_t} \mathbb{P}{\mbf \nu} dL \label{bl26.2}
\end{equation}
for $L_t$ (with normal $\mbf \nu$) and $l_t = {\mbf \chi}(L_t)$ (with normal $\bar{\mbf \nu}$) as curves on the referential and spatial singular surface, respectively. Let $\{{\mbf \nu}, {\mathbbm t}, {\mathbb N}\}$ be a positively oriented orthonormal basis on $S_r$. Define $\bar{\mathbbm t} \in T_{s_t}$ by $\bar{\mathbbm t} dl = {\bf F}^\pm {\mathbbm t}$. The triad $\{\bar{\mbf \nu}, \bar{\mathbbm t}, {\mathbbm n}\}$ then forms a positively oriented orthonormal basis on $s_t$, where $\mathbbm n$ is given by \eqref{sd17}$_1$. Hence $\bar{\mbf \nu} dl =  {j}^{-1} ({\bf F}^\pm {\mathbbm t} \times ({\bf
F}^\pm)^* {\mathbb N})dL$, which on repeated use of the definition of cofactor simplifies to
\begin{equation}
\bar{\mbf \nu} dl = {j}  {\mathbb F}^{-T}{\mbf \nu}dL.
\end{equation}
Stresses $\mathbb P$ and $\mathbb T$ are therefore related as (compare with ${\bf P} = J_F {\bf T} {\bf F}^{-T}$, where ${\bf T} \in Lin$ is the bulk Cauchy Stress)
\begin{equation}
\mathbb{P} = j \mathbb{T} {\mathbb F}^{-T}. \label{bl26.3}
\end{equation}

The balance laws in the spatial configuration,  equivalent to \eqref{bl18}, \eqref{bl19}, \eqref{bl26}, and \eqref{bl26.1}, are given by
\begin{eqnarray}
&& \rho \dot{\bf v} = \div{\bf T} + \rho {\bf b},~  {\bf T} = {\bf T}^T ~  \forall {\bf x} \in \kappa_t \setminus s_t, \label{bl26.39}
\\
&& j_s \llbracket{\bf v}\rrbracket + \llbracket{\bf T}\rrbracket{\mathbbm n} +\div^S \mathbb{T}  = {\bf 0}, ~\text{and}~ {\mathbb T} = {\mathbb T}^T   ~ \forall {\bf x} \in s_t, \label{bl26.4}
\end{eqnarray}
where $\rho$ is the mass density with respect to $\kappa_t$ and $j_s = {\rho_r U}j^{-1}$.

\paragraph{Dissipation inequalities} Let $\Psi$ and $\Phi$ be the free energy densities per unit volume of $\kappa_r$ and per unit area of $S_r$, respectively. Assume that $S_r$ has zero body force and kinetic energy density. For an arbitrary volume $\Omega \subset \kappa_r$, with $S = \Omega \cap S_r$ nonempty and $\partial S \subset \partial \Omega$, the mechanical version of second law of thermodynamics (under isothermal conditions) yields
\begin{equation}
\int_\Omega \rho_r{\bf b} \cdot {\bf v} dV + \int_{\partial \Omega} {\bf P N} \cdot{\bf v} dA +
\int_{\partial S} {\bf \mathbb{P}}{\mbf \nu} \cdot{\mathbbm v} dL
- \frac{d}{dt} \int_\Omega \left(\Psi + \frac{1}{2} \rho_r \left\vert {\bf v} \right\vert^2 \right)  dV - \frac{d}{dt} \int_{S} \Phi dA
\geq 0. \label{bl30}
\end{equation}

A comment is in order for the term representing the power due to interfacial stress. At every point on the curve $\partial S$ the contact force (between the surfaces divided by the curve) is given by ${\bf \mathbb{P}}{\mbf \nu}$ and the rate of change in displacement, with respect to an observer sitting on $S$ (at the considered point) and moving with velocity $U\mathbb{N}$, is given by ${\mathbbm v}$. The change in displacement apparent to the observer sitting on $S$ but moving tangentially to the interface will depend on the chosen parametrization and so will the resulting power. This is undesirable and therefore we use only ${\mathbbm v}$ to calculate the power expended at the interface. Gurtin and coauthors \cite{CermelliGurtin94, Gurtin00, GurtinStruthers90} have imposed invariance with respect to tangential velocities in their formulation of configurational balance laws. This is equivalent to the requirement of invariance under
re-parameterizations of the interface. Our viewpoint is different: We require ({\it a priori}) the mechanical power balance to be invariant under re-parametrization and write it in a form that satisfies this invariance
automatically. Thus this requirement is automatically satisfied in the
present formulation and accordingly yields no non-trivial information.

Before we proceed let us clarify the nature of interfacial stresses. The interface stress $\mathbb{P}$, in contrast to the bulk stress, does not act on a fixed set of material points but rather on material
points momentarily occupying the surface $S$. This is in accord with the mechanism responsible for surface tension in liquids. As the surface area increases, interstices are generated which are filled by molecules from the bulk liquid. In this way the surface tension remains sensibly constant while the surface area expands. Thus the matter occupying the surface does not actually stretch. Instead, the surface changes its area due to the continuous addition of mass. This physical situation stands in contrast to the treatment of surface tension in conventional continuum mechanics, in which the surface is regarded as a material surface if the motion of the liquid, regarded as a closed set, is continuous. In the conventional interpretation, surface tension is then a conventional force system acting on a persistent set of material points. However, in the actual physical situation, the surface is not material in the usual sense. Our framework accommodates such mechanisms while retaining the conventional interpretation of force. The contribution to mechanical power from interface stresses (as in \eqref{bl30} above) is consequently obtained not by its action on material velocities but on $\mathbbm{v}$.

Using the transport, divergence, and localization theorems, \eqref{bl30} reduces to
\begin{eqnarray}
&& {\bf P}\cdot \dot{\bf F} - \dot{\Psi} \geq 0  ~  \forall {\bf X} \in \kappa_r \setminus S_r ~ \text{and} \label{di1}
\\
&& \left \llbracket  {\bf P} {\mathbb N}\cdot {\bf v} + U \left(\Psi + \frac{1}{2} \rho_r \left\vert {\bf v} \right\vert^2 \right)  \right \rrbracket + \Div^S {\mathbb P}^T{\mathbbm v} - (\mathring{\Phi} - 2UH \Phi) \geq 0 ~  \forall {\bf X} \in S_r, \label{bl32}
\end{eqnarray}
where, in obtaining \eqref{di1}, we have also used balances of mass and momentum. We now rewrite \eqref{bl32} using the identities
\begin{eqnarray}
&& \llbracket  {\bf P} {\mathbb N}\cdot {\bf v} \rrbracket = \llbracket  {\bf P}  \rrbracket {\mathbb N}\cdot {\mathbbm v} - U  \llbracket  {\bf P} {\mathbb N}\cdot {\bf F}{\mathbb N} \rrbracket, \label{iden1}
\\
&&
\llbracket  {\bf v}\cdot {\bf v} \rrbracket = 2 \llbracket  {\bf v} \rrbracket \cdot \mathbbm{v} + U^2 \llbracket  | {\bf F}{\mathbb N} |^2 \rrbracket,  ~\text{and} \label{iden2}
\\
&&
\Div^S {\mathbb P}^T{\mathbbm v} = \Div^S {\mathbb P} \cdot {\mathbbm v} + {\mathbb P} \cdot \nabla^S{\mathbbm v}. \label{iden3}
\end{eqnarray}
Here \eqref{iden1} and \eqref{iden2} can be verified with the help of \eqref{s0.2}, \eqref{cc13}, and \eqref{sd7}$_2$ while \eqref{iden3} follows from the chain rule of differentiation. These identities, in addition to \eqref{bl4.4} and \eqref{bl19}, reduce \eqref{bl32} to
\begin{equation}
U \left({\mathbb N} \cdot \llbracket {\bf E} \rrbracket {\mathbb N} + \frac{1}{2} U^2 \rho_r \llbracket | {\bf F}{\mathbb N} |^2 \rrbracket \right) + {\mathbb P} \cdot \nabla^S{\mathbbm v} - (\mathring{\Phi} - 2UH \Phi ) \geq 0 ~  \forall {\bf X} \in S_r \label{bl}
\end{equation}
or equivalently (on substituting $\nabla^S{\mathbbm v}$ from \eqref{sd8.4}
and $2H = \tr \mathbb{L}$)
\begin{equation}
U \left({\mathbb N} \cdot \llbracket {\bf E} \rrbracket {\mathbb N} +
\frac{1}{2} U^2 \rho_r \llbracket | {\bf F}{\mathbb N} |^2 \rrbracket
\right) + U \mathbb{E} \cdot \mathbb{L}+ {\mathbb P} \cdot
\mathring{\mathbb F} \mathbbm{1} - \mathring{\Phi} \geq 0 ~  \forall {\bf
X} \in S_r, \label{di2}
\end{equation}
where
\begin{eqnarray}
&& {\bf E} = \Psi {\bf 1} - {\bf F}^T{\bf P} ~\text{and} \label{cdi23}
\\
&& {\mathbb E} = \Phi {\mathbbm 1} - {\mathbb F}^T{\mathbb P}
\label{cdi23.1}
\end{eqnarray}
are bulk and interface Eshelby tensors defined over $\kappa_r  \setminus S_r$ and $S_r$, respectively. Dissipation inequalities \eqref{di1} and \eqref{di2}, in addition to balance laws for mass and momentum, should be satisfied for every process.

\begin{rem} The present setting differs from that of Gurtin \cite{CermelliGurtin94, GurtinStruthers90, Gurtin00} as we do not consider any explicit contribution from configurational forces in the global dissipation inequality \eqref{bl30} (compare with Equations (21-6) and (21-19) in \cite{Gurtin00}); the final results however coincide. We demonstrate this by assuming, for now, the interface energy density to be of the form $\Phi = \hat{\Phi}({\mathbb F}, {\mathbb N})$. Such energies have been well studied in the contexts of phase equilibrium with interfacial energy \cite{GurtinStruthers90, LeoSekerka89, Silhavy11}. The surface stress is given by ${\mathbb P} = \hat{\Phi}_{\mathbb F} \mathbbm{1}$. We can then obtain $\mathring{\Phi} = {\mathbb P} \cdot \mathring{\mathbb F} \mathbbm{1} - \mathbbm{c} \cdot \mathring{\mathbb N}$, where $\mathbbm{c} = {\mathbb F}^T \hat{\Phi}_{\mathbb F} {\mathbb N} - \hat{\Phi}_{\mathbb N}$ is tangential. Substituting $\mathring{\mathbb N}$ from \eqref{sd5.1} and using the chain rule of differentiation yields
\begin{equation}
\mathring{\Phi} = {\mathbb P} \cdot \mathring{\mathbb F} \mathbbm{1}  - U \Div^S \mathbbm{c} + \Div^S (U\mathbbm{c}). \label{di2.002}
\end{equation}
Substituting it in \eqref{di2} we get
\begin{equation}
U \left({\mathbb N} \cdot \llbracket {\bf E} \rrbracket {\mathbb N} +
\frac{1}{2} U^2 \rho_r \llbracket | {\bf F}{\mathbb N} |^2 \rrbracket
\right) + U \mathbb{E} \cdot \mathbb{L}  + U \Div^S \mathbbm{c} - \Div^S (U\mathbbm{c}) \geq 0 ~  \forall {\bf
X} \in S_r. \label{di2.01}
\end{equation}
The term $\Div^S (U\mathbbm{c})$ drops out of the inequality. Indeed after integrating \eqref{di2.01} over $S$ and applying surface divergence theorem \eqref{st6.1} this term takes the form
\begin{equation}
\int_{\partial S} U {\mathbbm c} \cdot {\mbf \nu} dL, \label{di2.001}
\end{equation}
where ${\mbf \nu} \in T_{S_r}$ is the exterior unit normal to $\partial S$. Since $\Omega$ is arbitrary and $\mathbbm c$ is tangential, we can choose $\partial \Omega$ such that ${\mathbbm c} \cdot {\mbf \nu} = 0$ (i.e., orient $\partial \Omega$ such that $\mathbbm{c}$ is parallel to the tangent at every point on $\partial S$, cf. Figure \ref{intersect}). Upon localization of the resulting integral inequality we are finally led to
\begin{equation}
U \left({\mathbb N} \cdot \llbracket {\bf E} \rrbracket {\mathbb N} +
\frac{1}{2} U^2 \rho_r \llbracket | {\bf F}{\mathbb N} |^2 \rrbracket
 + \mathbb{E} \cdot \mathbb{L}  +  \Div^S \mathbbm{c} \right) \geq 0 ~  \forall {\bf
X} \in S_r \label{di2.1}
\end{equation}
as a necessary condition for \eqref{bl30}. The coefficient of $U$ is the net driving force for the motion of a coherent interface between two bulks phases. This coincides with the result obtained by Gurtin, cf. Equations (21-10a) and (21-26) in \cite{Gurtin00}. The configurational shear ${\mbf \tau}$ appearing in those equations from \cite{Gurtin00} is equal to ${\mathbbm c}$ (see $\S 2$ of \cite{Silhavy11} in this regard). At thermodynamic equilibrium the driving force vanishes thereby furnishing a balance relation to be satisfied at the interface. Such relations were also obtained via energy minimization \cite{JohnsonAlexander86, LarcheCahn78, LarcheCahn85, LeoSekerka89}. Ours is a dynamical theory, whereas results
coming from energy minimization are really only relevant at equilibrium,
and even then only for stable equilibria. \label{intcond}
\end{rem}


\subsection{\label{epd} Elastic plastic deformation}

The idea of stress-free local configurations is central to our theory. We assume both the bulk and the interface stress to be purely elastic in origin, wherein the deformation is measured with respect to the stress-free configuration. In a recent paper \cite{Guptaetal07} we demonstrated, using the mean-stress theorem, that it is always possible to obtain a locally stress-free state (under equilibrium and in the absence of external forces) by cutting $\kappa_t$ into parts with arbitrarily small volume. Moreover, if these sub-bodies cannot be made congruent in absence of any distortion then they do not form a connected set in a Euclidean space. The material is then said to be dislocated with no global differentiable map from $\kappa_t$ to the disjoint set of sub-bodies \cite{Bilbyetal55, Kondo55, Kondo64, Kroner81, Noll67}. The union of these unstressed sub-bodies is a three-dimensional non-Euclidean smooth manifold, say $\mathcal M$. A local configuration in $\mathcal{M}$ is identified with the local tangent space, denoted by $\kappa_i$. The local map from $\kappa_i$ to $\mathcal{V}_{\kappa_t}$ is represented by ${\bf H} \in Lin$. The absence of a global differentiable map renders $\bf H$ incompatible and therefore, unlike $\bf F$, it cannot be written as gradient of a differentiable map. The incompatibility of $\bf H$ implies the existence of a continuous distribution of dislocations over $\kappa_t$ (see the next section for details).

The argument used for the existence of stress-free local configurations in \cite{Guptaetal07} assumes smoothness of bulk stress. If the stress field is non-smooth only over a set of measure zero, the stresses can still be relaxed on neighborhoods arbitrarily close to the singular region and therefore everywhere except over the set of zero measure. If singular regions have stresses associated with them, for example the surface stresses discussed above, then they also need to be relaxed. In the following we show that this can be done under equilibrium and vanishing external forces if the surface $s_t$ is cut into infinitesimal areas.

To this end consider an arbitrary subsurface $s \subset s_t$ and assume $\mathbb{T}$ to be continuously differentiable over $s_t$. A simple calculation (using \eqref{bl26.4}$_1$ without the inertial term) then yields
\begin{equation}
\bar{\mathbb T} = \frac{1}{a}\int_{s}{\mathbb T} da = \frac{1}{2a} \left\{\int_{s} \rho({\bf x}\otimes \llbracket{\bf T}\rrbracket{\mathbbm n} + \llbracket{\bf T}\rrbracket{\mathbbm n} \otimes {\bf x}) da + \int_{\partial s} \rho ({\bf x}\otimes {\mathbb T}\bar{\mbf \nu} + {\mathbb T}\bar{\mbf \nu} \otimes {\bf x}) dl \right\},
\end{equation}
where $\bar{\mathbb T}$ is the mean interface Cauchy stress and $a$ is the area of $s$. The mean stress therefore vanishes if there are no external forces on $s$. According to the mean value theorem, there exists $\bar{\bf
x} \in s$ such that ${\mathbb T}(\bar{\bf x},t)=\bar{\mathbb T}$ ($={\bf 0}$). Let the area $a$ become arbitrarily close to zero. Then, by continuity of $\mathbb T$, the surface stress $\mathbb T$ can be brought arbitrarily close to zero.

\begin{figure}[h!]
\centering
\includegraphics[width=4.6in]{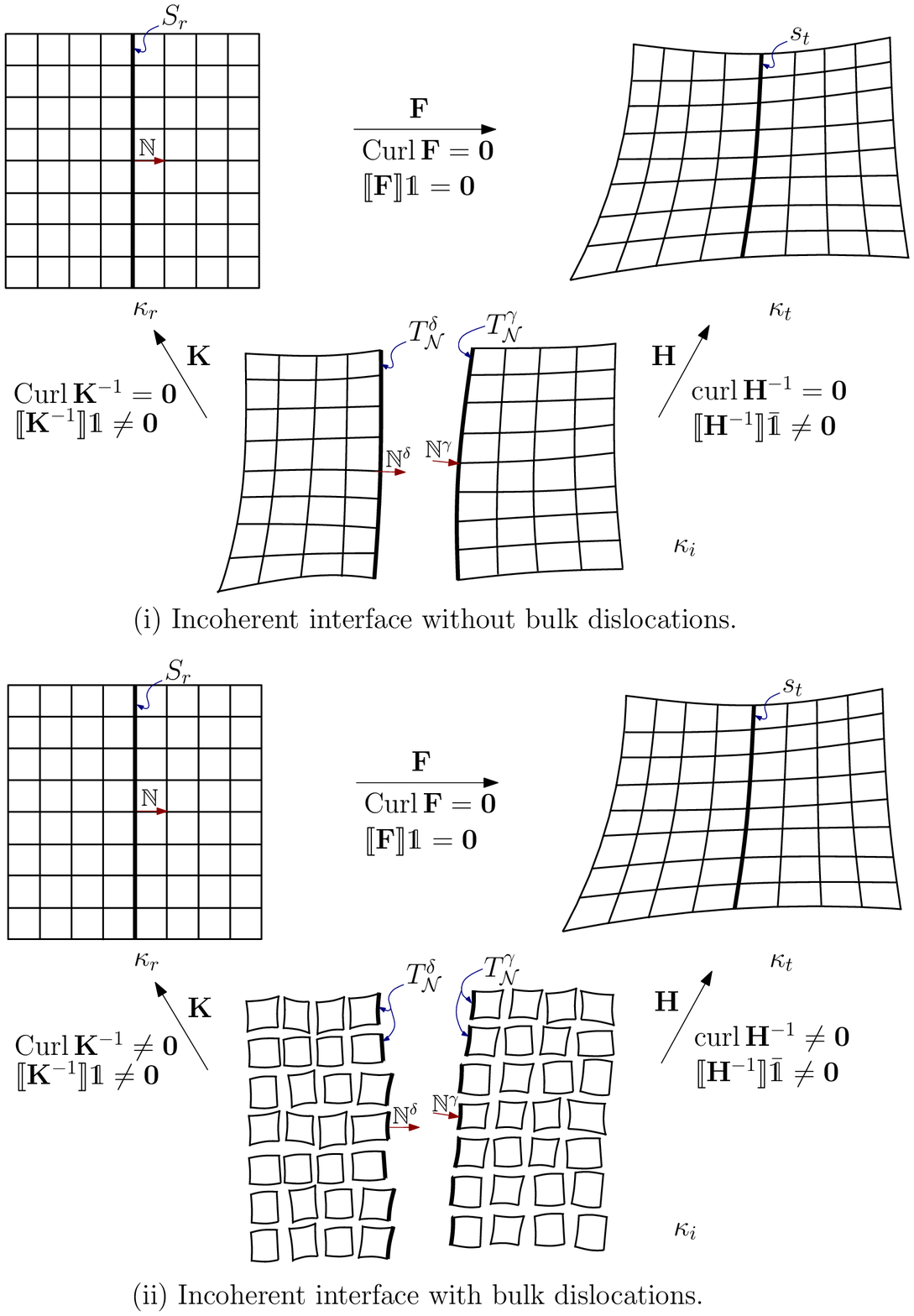}
\caption{A cartoon illustrating relaxed configurations. The rectangular blocks represent infinitesimal neighborhood in the body. In case (i) the bulk is free of defects and therefore the relaxed manifold consists of two disjoint global configurations which do not fit together at the interface. When the bulk is dislocated, as in case (ii), the relaxed manifold is a collection of only local configurations which do not fit together with each other.}
\label{relax}
\end{figure}

While cutting $\kappa_t$, care is needed with surfaces where the bulk stress is singular. The neighborhood of a point on such surfaces is to be cut such that the length dimension parallel to the normal (of the surface) is arbitrarily small compare to other length dimensions. This way we will be left essentially with areas to be relaxed from stress, if any. The resulting stress-free configurations at the singular interface are of dimension one less than those obtained from the bulk. Their union forms a two dimensional smooth manifold $\mathcal{N}$. A local configuration in $\mathcal{N}$ is identified with the local tangent space of $\mathcal{N}$. If the tangent space $T_{s_t}({\bf x})$ is  mapped (locally) into two disjoint local configurations in $\mathcal{N}$, for reasons that will become clear below, then we call the singular interface incoherent (at $\bf x$). We denote the two local configurations by $T_\mathcal{N}^\gamma$ and $T_\mathcal{N}^\delta$ (in rest of the paper, a superscript $\gamma$ will represent an association with $T_\mathcal{N}^\gamma$ configuration and $\delta$ with $T_\mathcal{N}^\delta$; they are not to be confused as indices). Otherwise, if the mapping is injective then we call the singular interface coherent and denote the local configuration by $T_\mathcal{N}$. Incoherency of the interface implies a continuous distribution of dislocations over the interface; we postpone the discussion on this aspect till the next Section. The process of relaxation is illustrated through a cartoon in Figure \ref{relax}.

Let ${\bf K} \in Lin$ be the local map from tangent space $\kappa_i$ to $\mathcal{V}_{\kappa_r}$ at ${\bf X} \in \kappa_{r} \setminus S_r$. Both $\bf H$ and $\bf K$ are assumed to be continuously differentiable except on the singular surface. The following decomposition
\begin{equation}
{\bf H}={\bf FK} ~~ \forall {\bf X} \in \kappa_{r} \setminus S_r \label{dai2}
\end{equation}
is admitted (conventional plasticity theories usually represent tensors ${\bf H}$ and ${\bf K}^{-1}$ by ${\bf F}^e$ and ${\bf F}^p$, respectively). Since we demand unloading to be elastic in nature, we call $\bf H$ the elastic distortion. We identify $\bf K$ with plastic distortion, for reasons that will become apparent when we discuss dissipation in Subsection \ref{pidi}. Define distortion maps on the surface
\begin{eqnarray}
&& \mathbb{H}^\gamma = {\bf H}^+ \mathbbm{1}^\gamma, ~\mathbb{H}^\delta = {\bf H}^- \mathbbm{1}^\delta~\text{and} \label{epd1}
\\
&& \mathbb{K}^\gamma = {\bf K}^+ \mathbbm{1}^\gamma, ~\mathbb{K}^\delta = {\bf K}^- \mathbbm{1}^\delta, \label{epd2}
\end{eqnarray}
where superscripts $\gamma$ and $\delta$ denote the association with the two local configurations in $\mathcal{N}$ at a fixed material point. The projection tensors $\mathbbm{1}^\gamma: \mathcal{V} \rightarrow T_\mathcal{N}^\gamma$ and $\mathbbm{1}^\delta: \mathcal{V} \rightarrow T_\mathcal{N}^\delta$  are given by
\begin{equation}
\mathbbm{1}^\gamma = {\bf 1} - {\mathbb N}^\gamma \otimes {\mathbb N}^\gamma~\text{and}~\mathbbm{1}^\delta = {\bf 1} - {\mathbb N}^\delta \otimes {\mathbb N}^\delta, \label{epd3}
\end{equation}
where ${\mathbb N}^\gamma$ and ${\mathbb N}^\delta$ are unit normals to $T_\mathcal{N}^\gamma$ and $T_\mathcal{N}^\delta$, respectively. They are related to ${\mathbb N}$ and $\mathbbm n$ as (cf. \eqref{sd17}$_1$)
\begin{eqnarray}
&& {\mathbbm n} = \frac{({\bf H}^\pm)^{-T} {\mathbb N}^\alpha}{|({\bf H}^\pm)^{-T} {\mathbb N}^\alpha|}~\text{and} \label{epd4}
\\
&& {\mathbb N} = \frac{({\bf K}^\pm)^{-T} {\mathbb N}^\alpha}{|({\bf K}^\pm)^{-T} {\mathbb N}^\alpha|}, \label{epd5}
\end{eqnarray}
where $\alpha = \{\gamma, \delta\}$. Here, superscript $+$ appears with $\gamma$ and $-$ with $\delta$ (either pair can be equivalently used to obtain ${\mathbbm n}$ or $\mathbb{N}$). Normals ${\mathbb N}^\gamma$ and ${\mathbb N}^\delta$ will coincide only if the jump $\llbracket {\bf H}^{-1}\rrbracket$ (or $\llbracket {\bf K}^{-1}\rrbracket$) across the interface is of Hadamard's rank one form, i.e. if $\llbracket {\bf H}^{-1}\rrbracket = {\bf h} \otimes {\mathbbm n}$ (or $\llbracket {\bf K}^{-1}\rrbracket = {\bf k} \otimes \mathbb{N}$ for arbitrary $\bf h$ and $\bf k$). Otherwise they will be distinct, resulting in two distinct local configurations after stress relaxation at each material point on the interface. The former case leads to a coherent interface and the latter to an incoherent interface. Let $j^\alpha$ ($j^\alpha_s$) be the ratios of infinitesimal area on $S_r$ ($s_t$) to local infinitesimal areas on $T_\mathcal{N}^\alpha$. They are given by
\begin{equation}
j^\alpha = |({\bf K}^\pm)^{*} {\mathbb N}^\alpha|~\text{and}~j^\alpha_s = |({\bf H}^\pm)^{*} {\mathbb N}^\alpha| \label{epd6}
\end{equation}
and are related to each other as
\begin{equation}
 j^\alpha_s = j \left( j^\alpha\right)^{-1}, \label{epd6.1}
 \end{equation}
where $j$ is the ratio of infinitesimal area on $s_t$ to $S_r$.

A straightforward calculation, using \eqref{epd4} and \eqref{epd5}, confirms that
\begin{equation}
\mathbb{H}^\alpha = \bar{\mathbbm{1}} \mathbb{H}^\alpha,~\text{and} ~\mathbb{K}^\alpha = {\mathbbm{1}} \mathbb{K}^\alpha. \label{epd7}
\end{equation}
Therefore $\mathbb{H}^\alpha$ map local configurations in $\mathcal{N}$ to $T_{s_t}$ and $\mathbb{K}^\alpha$ map local configurations in $\mathcal{N}$ to $T_{S_r}$. Employing \eqref{epd1}, \eqref{epd2}, \eqref{epd7}, and \eqref{sd7}$_1$, and taking limiting values of \eqref{dai2}, we are led to the following multiplicative decompositions on the singular surface: (compare with \eqref{dai2})
\begin{equation}
\mathbb{H}^\alpha = \mathbb{F} \mathbb{K}^\alpha ~\forall {\bf X} \in S_r, ~\text{with}~\alpha = \{\gamma, \delta\}. \label{epd8}
\end{equation}

There exist unique pseudoinverse tensors $\left(\mathbb{H}^\alpha\right)^{-1}$ and $\left(\mathbb{K}^\alpha\right)^{-1}$ such that (here and elsewhere, no summation is implied for repeated superscript $\alpha$ unless explicitly stated)
\begin{eqnarray}
&& \left(\mathbb{H}^\alpha\right)^{-1} \mathbb{H}^\alpha = {\mathbbm 1}^\alpha, ~  \mathbb{H}^\alpha \left(\mathbb{H}^\alpha\right)^{-1} = \bar{\mathbbm 1}, \label{epd9}
\\
&& \left(\mathbb{K}^\alpha\right)^{-1} \mathbb{K}^\alpha = {\mathbbm 1}^\alpha, ~\text{and}~ \mathbb{K}^\alpha \left(\mathbb{K}^\alpha\right)^{-1} = {\mathbbm 1}. \label{epd10}
\end{eqnarray}
Identities
\begin{equation}
\left(\mathbb{H}^\alpha\right)^{-1} = \left({\bf H}^\pm\right)^{-1} \bar{\mathbbm{1}} ~\text{and}~ \left(\mathbb{K}^\alpha\right)^{-1} = \left({\bf K}^\pm\right)^{-1} \mathbbm{1} \label{epd11}
\end{equation}
are then immediate, as can be verified by direct substitution using \eqref{epd1}, \eqref{epd2}, and \eqref{epd7}. Here superscript $+$ is used to define $\left({\mathbb H}^\gamma\right)^{-1}$ (and $\left(\mathbb{K}^\gamma\right)^{-1}$) and $-$ for $\left(\mathbb{H}^\delta\right)^{-1}$ (and $\left(\mathbb{K}^\delta\right)^{-1}$).

Finally, on the basis of our earlier discussion regarding $\mathbb{F}$, we can obtain the following results for interfacial elastic and plastic distortions: (compare with \eqref{sd8.1}, \eqref{deronsur4}, and \eqref{deronsur5})
\begin{eqnarray}
&& \left(\mathbb{K}^{\alpha}\right)^\ast = {j}^\alpha ({\mathbb N} \otimes \mathbb{N}^\alpha),~ j^\alpha_{\mathbb{K}^\alpha}=j^\alpha (\mathbb{K}^\alpha)^{-T},~j^\alpha_{\mathbb{N}^\alpha} = {\bf 0},~ \mathring{j^\alpha} =  j^\alpha \mathring{\mathbb{K}}^\alpha({\mathbb{K}^\alpha})^{-1} \cdot {\mathbbm 1}, \label{epd12}
\\
&& \left(\mathbb{H}^{\alpha}\right)^\ast = {j}_s^\alpha ({\mathbbm n} \otimes \mathbb{N}^\alpha),~ ({j_s^\alpha})_{\mathbb{H}^\alpha}={j}_s^\alpha (\mathbb{H}^\alpha)^{-T},~({j_s^\alpha})_{\mathbb{N}^\alpha} = {\bf 0},~\text{and}~ \mathring{{j}_s^\alpha} =  {j}_s^\alpha \mathring{\mathbb{H}}^\alpha({\mathbb{H}^\alpha})^{-1} \cdot \bar{\mathbbm 1} \label{epd13}
\end{eqnarray}
for each $\alpha = \{\gamma, \delta\}$.

\subsection{\label{ssm}Symmetries of space and matter} To develop physically consistent constitutive models we need to exploit spatial and material symmetries as afforded by the structure of the space and the material. We will obtain restrictions on the form of constitutive response functions (for example stress, energy, kinetic laws), both in bulk and on interface, upon imposing their invariance under the symmetries. In this subsection we discuss kinematical changes induced by the symmetries but postpone their application to response functions until Section \ref{ipdkl}.

\paragraph{Material frame indifference} The material response is expected to remain invariant under arbitrary changes in the frame of reference. For the case at hand this amounts to requiring invariance under superimposed rigid body rotations. The motion transforms as ${\bf x} \rightarrow {\bf Q}(t) {\bf x} + {\bf c}(t)$, where ${\bf Q} \in Orth^+$ and ${\bf c} \in \mathcal{V}$ are functions of time only (arrow denotes the transformation), and the deformation gradient as ${\bf F} \rightarrow {\bf Q} {\bf F}$. We assume ${\bf T} \rightarrow {\bf QT}{\bf Q}^{T}$ for Cauchy stress and therefore ${\bf P}\rightarrow {\bf QP}$.

The transformations for $\bf H$ and $\bf K$, obtained by utilizing the constitutive framework of elastic plastic deformation in \cite{Guptaetal07}, are given by ${\bf H} \rightarrow {\bf Q}{\bf H}$ and ${\bf K}\rightarrow {\bf K}$; while for unit normal ${\mathbbm n}$ it follows from \eqref{sd17}$_1$, ${\mathbbm n} \rightarrow {\bf Q} {\mathbbm n}$. All other unit normals: ${\mathbb N}$ and ${\mathbb N}^\alpha$ are invariant, where $\alpha = \{\gamma, \delta\}$. As a consequence $\bar{\mathbbm{1}} \rightarrow {\bf Q}\bar{\mathbbm{1}} {\bf Q}^T$, $\mathbbm{1} \rightarrow \mathbbm{1}$, and $\mathbbm{1}^\alpha \rightarrow \mathbbm{1}^\alpha$. Using these with  \eqref{epd1} and \eqref{epd2}, we obtain $\mathbb{H}^\alpha \rightarrow {\bf Q} \mathbb{H}^\alpha$ and $\mathbb{K}^\alpha \rightarrow  \mathbb{K}^\alpha$. Moreover, $\left(\mathbb{H}^\alpha\right)^{-1} \rightarrow \left(\mathbb{H}^\alpha\right)^{-1}{\bf Q}^T $ and $\left(\mathbb{K}^\alpha\right)^{-1} \rightarrow  \left(\mathbb{K}^\alpha\right)^{-1}$, cf. \eqref{epd11}.

\paragraph{Compatible changes in the reference configuration} Our choice of reference configuration is arbitrary as long as it is a connected subset of the Euclidean space. We define compatible changes in the reference configuration as those maps which preserve its connectedness and Euclidean nature. A constitutive response function should be invariant with respect to arbitrary compatible changes in the reference configuration unless it is defined explicitly with respect to a particular reference configuration (which is the case in conventional elasticity theories, where the reference configuration is usually the stress-free configuration) \cite{Davini88, EpsteinElz07, Guptaetal07}.

At a fixed time consider a variation of $\kappa_{r}$ from $\kappa_{r_{1}}$ to $\kappa_{r_{2}}$ defined by a continuous but piecewise differentiable bijective map ${\bf X}_{2}={\mbf \lambda} ({\bf X}_{1})$, where ${\bf X}_1 \in \kappa_{r_1}$ and ${\bf X}_2 \in \kappa_{r_2}$, with invertible gradient ${\bf A}=\nabla
_{1}{\mbf \lambda}$, where $\nabla_{1}$ is the gradient with respect to ${\bf X}_{1}$. To maintain compatibility of $\mbf \lambda$ at the singular surface, tensor $\bf A$ can at most have Hadamard's rank one jump discontinuity at the interface, i.e. $\llbracket{\bf A}\rrbracket = {\bf d}\otimes{\mathbb N}_1$ for arbitrary ${\bf d}\in \mathcal{V}$. Local configurations in $\mathcal{M}$ and $\mathcal{N}$, and the global spatial configuration $\kappa_t$, all remain invariant under compatible changes in $\kappa_r$. Using obvious notation we have ${\bf K}_{1}^{-1}d{\bf X}_{1}={\bf K}_{2}^{-1}d{\bf X}_{2}$ away from the interface, leading to
\begin{equation}
{\bf K}_2 = {\bf A}{\bf K}_1,~{\bf H}_2 = {\bf H}_1,~\text{and} ~{\bf F}_2 = {\bf F}_1{\bf A}^{-1}. \label{invcomp1}
\end{equation}
The unit normal ${\mathbb N}_1 \perp T_{S_{r_1}}$ at some ${\bf X}_1 \in S_{r_1}$ changes to
\begin{equation}
{\mathbb N}_{2}=j_A^{-1} \left({\bf A}^\pm\right)^{\ast}{\mathbb N}_{1}  \label{invcomp1.01}
\end{equation}
such that ${\mathbb N}_2 \perp T_{S_{r_2}}$ at ${\bf X}_2 = {\mbf \lambda}({\bf X}_1) \in S_{r_2}$, where $j_A$ is the ratio of infinitesimal areas on $S_{r_2}$ and $S_{r_1}$. Other unit normal fields, ${\mathbb N}^\alpha$ ($\alpha = \{\gamma, \delta\}$) and $\mathbbm n$ remain invariant. For interfacial distortions, recall \eqref{epd1}, \eqref{epd2}, and \eqref{epd7} to write
\begin{equation}
\mathbb{K}^\alpha_2 = \mathbb{A} \mathbb{K}^\alpha_1~\text{and}~ \mathbb{H}^\alpha_2 = \mathbb{H}^\alpha_1, \label{invcomp1.02}
\end{equation}
where $\mathbb{A} = {\bf A}^\pm \mathbbm{1}_1$. Moreover
\begin{equation}
\left(\mathbb{K}^\alpha\right)^{-1}_2 =  \left(\mathbb{K}^\alpha\right)^{-1}_1 \mathbb{A}^{-1},~ \left(\mathbb{H}^\alpha\right)^{-1}_2 = \left(\mathbb{H}^\alpha\right)^{-1}_1,~\text{and}~ {\mathbb F}_2 = {\mathbb F}_1 {\mathbb A}^{-1},  \label{invcomp1.03}
\end{equation}
where ${\mathbb A}^{-1}$ is the pseudoinverse of $\mathbb{A}$ such that ${\mathbb A}^{-1} {\mathbb A} = \mathbbm{1}_1$,
${\mathbb A} {\mathbb A}^{-1} = \mathbbm{1}_2$, and $\mathbb{A}^{-1} = \left({\bf A}^\pm\right)^{-1} \mathbbm{1}_2$, cf. \eqref{sd8.11} and \eqref{sd8.12}. Relations \eqref{invcomp1.03} can be verified by direct substitution using \eqref{epd9}, \eqref{epd10}, and \eqref{epd11}.

In fact, it turns out that $\bf A$ has to be continuous across the interface; i.e., ${\bf d} = {\bf 0}$. To show this we start by noting the change in $U_1$,
\begin{equation}
U_2 =  j_A^{-1} J_{A^\pm} U_1. \label{inv16}
\end{equation}
This follows upon recalling the discussion leading to \eqref{sd17} and using $|\left({\bf A}^\pm\right)^{-T} {\mathbb N}_1| = j_A J_{A^\pm}^{-1}$ from \eqref{invcomp1.01}. For $U_2$ to be single valued we require $J_{A^+} = J_{A^-}$. Applying this to the identity $J_{A^+} = J_{A^-} (1 + ({\bf A}^-)^{-1} {\bf d} \cdot {\mathbb N}_1)$, cf. Equation $(1.1.6)$ in \cite{Silhavy97}, we get ${\bf d} \cdot ({\bf A}^-)^{-T} \mathbb{N}_1 = 0$ i.e., ${\bf d} \in T_{S_{r_2}}$. A similar argument yields ${\bf e} \in T_{S_{r_1}}$, where $\bf e$ satisfy $\llbracket {\bf A}^{-1} \rrbracket = {\bf e} \otimes {\mathbb N}_2$. Next, consider an arbitrary single valued scalar field, $\pi$, defined on $S_{r_1}$. Let $\pi$ be invariant under the transformation. Denote the normal time derivative of $\pi$ with respect to $S_{r_2}$ by $\overset{\diamond}{\pi}$. Use \eqref{invcomp1.01}, \eqref{inv16}, and the definition of normal time derivative to get $\overset{\diamond}{\pi} = \mathring{\pi} + U_2 \nabla^S_1 \pi \cdot \mathbbm{1}_1 ({\bf A}^\pm)^{-1} {\mathbb N}_2$. Subtract these two equations (one with superscript $+$ and the other with superscript $-$) from each other, and use the continuity of $\overset{\diamond}{\pi}$, $\mathring{\pi}$, and $ U_2 \nabla^S_1 \pi$, to deduce $\mathbbm{1}_1 {\bf e} = {\bf 0}$ or, since ${\bf e} \in T_{S_{r_1}}$, ${\bf e} = {\bf 0}$. Therefore, ${\bf A}^{-1}$ (or equivalently $\bf A$) is continuous across the singular interface.

We need to make two assumptions regarding the nature of $\bf A$ at the interface. We assume the normal time derivative to remain invariant i.e.,  $\overset{\diamond}{\pi} = \mathring{\pi}$ for arbitrary $\pi$ and $\mathring{\mathbb A} = {\bf 0}$. Both of these are motivated by the requirement of dissipation, given in \eqref{tdi17} below, to remain invariant under the transformation in reference configuration. The former of these assumptions leads to $\mathbbm{1}_1 {\bf A}^{-1} {\mathbb N}_2 = {\bf 0}$ implying
\begin{equation}
{\bf A}^{-1} = h {\mathbb N}_1 \otimes {\mathbb N}_2 + {\mathbb A}^{-1} \label{invcomp1.011}
\end{equation}
at the interface, where $h = j_A J_A^{-1}$. The second assumption, on using the definition of normal time derivative, yields
\begin{equation}
\nabla_1 {\bf A} {\mathbb N}_1 = {\bf 0}. \label{invcomp1.012}
\end{equation}

Finally, we evaluate the transformation of the curvature tensor $\mathbb{L}$. Use \eqref{invcomp1.011} to rewrite \eqref{invcomp1.01} as
\begin{equation}
{\mathbb N}_{2}=h {\bf A}{\mathbb N}_{1}.  \label{invcomp1.013}
\end{equation}
Differentiate it and apply \eqref{invcomp1.012}, after noting the symmetry in the gradient of $\bf A$, to get
\begin{equation}
\mathbb{L}_2 = h \mathbb{A} \mathbb{L}_1 \mathbb{A}^{-1}. \label{inv18.1}
\end{equation}

\paragraph{Material symmetry} The concept of material symmetry is related to local configurations in the relaxed manifold $\mathcal{M}$. Let $\mathcal{G}$ be the symmetry group associated with local configuration $\kappa_i$. An element ${\bf G} \in \mathcal{G}$ then brings about a local change in $\kappa_i$, modifying distortions from $\bf H$ and $\bf K$ to ${\bf HG}$ and ${\bf KG}$, respectively, such that the constitutive response functions remain invariant. For solids, $\mathcal{G} \subseteq Orth^+$. Consider, for example, the strain energy density $W ({\bf H})$ (defined with respect to $\mathcal{M}$). Then, $W ({\bf H}) = W ({\bf HG})$. We assume the material to be uniform away from the interface thereby allowing $\bf G$ to be discontinuous at the interface. Note that the material neighborhoods separated by an interface can have distinct symmetry groups, as is generally the case with poly-crystalline materials.

The situation at the interface is more involved. For an incoherent interface we have two disjoint local configurations at each material point on $S_r$ and each of these can have a distinct symmetry group associated with it. Accordingly, we denote $\mathcal{G}^\gamma$ and $\mathcal{G}^\delta$ as distinct symmetry groups associated with local configurations $T_{\mathcal{N}}^\gamma$ and $T_{\mathcal N}^\delta$, respectively. Elements  ${\bf G}^\gamma \in \mathcal{G}^\gamma$ and ${\bf G}^\delta \in \mathcal{G}^\delta$ thereby modify distortions ${\bf H}^\pm$ and ${\bf K}^\pm$ to ${\bf H}^\pm{\bf G}^\alpha$ and ${\bf K}^\pm{\bf G}^\alpha$, with $\alpha = \gamma$ when the superscript is $+$ and $\delta$ when it is $-$ (no summation implied), such that constitutive response functions on the interface remain invariant. Assume ${\mathcal G}^\alpha \subseteq Orth^+$. This is sufficient to ensure that symmetry transformations neither strain the interface nor bring about a change in its local area, as shown below. Under the action of a symmetry map the interfacial normals ${\mathbb{N}}^\gamma$ and ${\mathbb N}^\delta$ are modified to $({\bf G}^\gamma)^T {\mathbb{N}}^\gamma$ and $({\bf G}^\delta)^T{\mathbb N}^\delta$, respectively. This follows immediately on using Nanson's formula and ${\bf G}^\alpha \in Orth^+$. As a result identity tensors $\mathbbm{1}^\alpha$ transform to $\mathbbm{1}^{\alpha'}=({\bf G}^\alpha)^T \mathbbm{1}^\alpha {\bf G}^\alpha$, where $\mathbbm{1}^{\alpha'}$ represent surface identity tensors for transformed local configurations at the interface. Using these we can obtain the transformation for interface distortion tensors, defined in \eqref{epd1} and \eqref{epd2}, as
\begin{equation}
{\mathbb H}^\alpha \rightarrow {\mathbb H}^\alpha \mathbb{G}^\alpha~\text{and}~{\mathbb H}^\alpha \rightarrow {\mathbb H}^\alpha \mathbb{G}^\alpha, \label{invcomp1.031}
\end{equation}
where $\mathbb{G}^\alpha = \mathbbm{1}^\alpha {\bf G}^\alpha$ (no summation over $\alpha$). Tensors $\mathbb{G}^\alpha$ satisfy
\begin{equation}
\mathbb{G}^\alpha (\mathbb{G}^\alpha)^T = \mathbbm{1}^\alpha,~ (\mathbb{G}^\alpha)^T \mathbb{G}^\alpha = \mathbbm{1}^{\alpha'},~\text{and}~|({\mathbb G}^\alpha)^\ast| = 1. \label{invcomp1.05}
\end{equation}
The last of these can be proved using $({\bf G}^\alpha)^\ast = {\bf G}^\alpha$ and $(\mathbbm{1}^\alpha)^\ast = {\mathbb N}^\alpha\otimes{\mathbb N}^\alpha$ (no summation on $\alpha$). Relations \eqref{invcomp1.05}$_{2,3}$ ensure that symmetry maps do not introduce any additional strain at the interface and bring about no change in its local area. The former of these claims follows upon noting the definition of interfacial strains,
\begin{equation}
{\mathbbm e}^\alpha = \frac{1}{2} \left({\mathbb C}^\alpha - \mathbbm{1}^\alpha \right), \label{invcomp1.04}
\end{equation}
where ${\mathbb C}^\alpha = \left({\mathbb H}^\alpha\right)^T {\mathbb H}^\alpha$. The latter claim is a consequence of the fact that $| ({\mathbb G}^\alpha)^\ast|$ represents the change in the area of local configurations under the action of symmetry map.

\begin{rem} (Continuous and discrete symmetry groups)
If the symmetry group $\mathcal{G}$ is continuous (for example isotropy or transversely isotropy) then the dislocation density $\mbf \alpha$ (defined in \eqref{dai6} below) fails to be a characteristic of a body (cf. Theorem $8$ in \cite{Noll67} and $\S 1.2.5$ in \cite{EpsteinElz07}). Indeed, at a fixed time for a material point away from the interface, any two intermediate configurations $\kappa_{i_1}$ and $\kappa_{i_2}$ are related by ${\bf K}_2 ({\bf X})= {\bf K}_1 ({\bf X}) {\bf G}({\bf X}) {\bf A}$ (cf. Theorem $2$ in \cite{Noll67}), where ${\bf G}({\bf X}) \in \mathcal{G}$ and ${\bf A} \in InvLin$ is constant. If the symmetry group is discrete then ${\bf K}_2 ({\bf X})$ can be a smooth field only if ${\bf G}$ is constant. However, for the continuous symmetry group, ${\bf G}$ can be a continuous function of $\bf X$. This leads to a uniquely defined dislocation density tensor only in the former case (for details see the cited references). It should be noted that for metals such continuous symmetries are, in practice, used to model a poly-crystalline material which, at a sufficient macroscopic scale, is considered as a random aggregate of single crystals. To model a poly-crystalline material within our framework, we would need to construct the theory at the level of an individual grain boundary, which separates single crystals. This is precisely one of the motivations for the current work.
\end{rem}

\section{Continuum distribution of dislocations: bulk and interface}

A distortion field ($\bf F$, $\bf H$, or $\bf K$) is said to be compatible, away from the singular interface, if it is given by the gradient of a differentiable vector field; and compatible at the interface if it admits a rank one jump of Hadamard's type across the interface. Within our model we assume both ${\bf H}$ and ${\bf K}$ to be  incompatible, and $\bf F$ to be compatible, everywhere in the body at all times. As we shall see below in Subsection \ref{idd}, incompatibility of  $\bf H$ (or equivalently of $\bf K$) in effect contributes to the Burgers vector for
arbitrary circuits in $\kappa_t$ (or $\kappa_r$) thus affording a relationship with continuous distributions of
dislocations. The vanishing of Burgers vector is equivalent to the compatibility of distortion fields $\bf H$ and $\bf K$ in the region. However, caution is required when interpreting a continuous distribution of dislocations in terms of a distribution of discrete dislocations. It is possible to have multiple arrangements of discrete dislocations which correspond to the same continuous dislocation density. This point, in the context of interface dislocation density, has been well illustrated in Section $2.4$ of \cite{SuttonBalluffi09}. Moreover, interface dislocation density can be represented by a two-dimensional array of discrete dislocations only for low magnitudes of dislocation density. This follows from the fact that individual dislocations cannot be placed arbitrarily close to each other (cf. Section $38$ in \cite{Christian02}). Similar remarks hold for bulk dislocation density.

The dislocation densities obtained in Subsection \ref{idd} are not invariant under compatible changes
in the reference configuration. This is unacceptable for a genuine measure of defect content which should be independent of the choice of a reference configuration. We are led to obtain invariant measures of bulk and interface dislocation densities which thus qualify as argument of constitutive response functions associated with plastic evolution, cf. \cite{EpsteinElz07, Guptaetal07} for bulk and Subsections \ref{pidi} and \ref{kr} for interface.

The dislocation densities associated with bulk and interface are not independent of each other. As shown in Subsection \ref{jbdd}, the projection of the jump in bulk dislocation density along the interface normal is related to the (surface) divergence of interface dislocation density. This relation immediately proves that interface dislocation density, in contrast to its bulk counterpart, does not have a vanishing divergence.

\subsection{\label{idd} Interface dislocation density}

Define Burgers vector
\begin{equation}
{\bf B}(C,t)=\int_C {\bf K}^{-1} d{\bf X}, \label{pwsl1}
\end{equation}
where $C$ is a close material curve which intersects $S_r$ in finite
number of points and ${\bf K}^{-1}$ is piecewise smooth. The plastic distortion is singular
on $S_r$ and therefore the integral in \eqref{pwsl1} will have
singularities only over a set of measure zero (a finite collection
of points on a line constitute such a set). Let $A_C$ be the area of
the surface enclosed by $C$ and let $\Gamma = A_C \cap S_r$ be the
curve which lies at the intersection of $A_C$ and $S_r$. Use the
Stokes' theorem for piecewise continuous tensor fields (see for example \cite{Guptaetal07})
to obtain
\begin{equation}
\int_C {\bf K}^{-1}d{\bf X} = \int_{A_C}(\Curl{\bf K}^{-1})^{T}{\bf
N}_C dA - \int_{\Gamma }\llbracket {\bf K}^{-1} \rrbracket d{\bf X},
\label{pwsl2}
\end{equation}
where ${\bf N}_C$ is the unit normal field associated with $A_C$. Let $\{{\mathbbm t}_{1},{\mathbbm t}_{2}\} \in T_{S_r}({\bf X})$ be such that
$\{{\mathbbm t}_{1},{\mathbbm t}_{2},{\mathbb N}\}$ is a positively oriented
orthogonal basis at ${\bf X} \in S_r$. Orient ${\mathbbm t}_2$ along
the curve $\Gamma$ so that $d{\bf X} = {\mathbbm t}_2 dL =({\mathbb N}
\times {\mathbbm t}_1)dL$ on $\Gamma$. Define a tensor field ${\mbf
\beta}_{r}$ on $S_r$, the (referential) interface dislocation
density, such that
\begin{equation}
\llbracket {\bf K}^{-1}\rrbracket ({\mathbbm t}_1 \times {\mathbb N})={\mbf
\beta} _{r}^{T} {\mathbbm t}_1. \label{pwsl4}
\end{equation}
Recognizing $\Curl{\bf K}^{-1}$ as the (referential) bulk dislocation density, denoted by ${\mbf \alpha}_{r}$, the net Burgers vector associated with $C$ can be written as a function of dislocation densities:
\begin{equation}
{\bf B}(C,t)=\int_{A_C} {\mbf \alpha}_{r}^{T} {\bf N}_CdA + \int_{\Gamma }
{\mbf \beta}_{r}^{T}{\mathbbm t}_1dL. \label{pwsl5}
\end{equation}
Writing $\llbracket {\bf K}^{-1}\rrbracket=\llbracket {\bf
K}^{-1}\rrbracket {\bf 1}$ with ${\bf 1} = {\bf N} \otimes {\bf N} +
{\bf t}_{a} \otimes {\bf t}_{a}$ ($a=1,2$), \eqref{pwsl4} leads to
\begin{equation}
\llbracket {\bf K}^{-1} \rrbracket = {\bf k} \otimes {\mathbb N} - {\mbf \beta}
_{r}^{T} {\mbf \varepsilon}_{({\mathbb N})}, \label{pwsl6}
\end{equation}
where ${\bf k} \in \mathcal{V}$ is arbitrary and
\begin{equation}
{\mbf \varepsilon}_{({\mathbb N})}={\mathbbm t}_{1}\otimes {\mathbbm t}_{2}-{\mathbbm t}_{2}\otimes {\mathbbm t}_{1} \label{pwsl7}
\end{equation}
is the two dimensional permutation tensor density on $T_{S_r}({\bf X})$. It
satisfies ${\mbf \varepsilon}_{({\mathbb N})}={\bf R} {\mbf \varepsilon}_{({\mathbb N})}{\bf R}^{T}$ for all
two dimensional orthogonal transformations $\bf R$ that preserve the orientation
of $T_{S_r}$ at a material point. Therefore, any pair of vectors in $T_{S_r}({\bf X})$ which with $\mathbb N$
form a positively oriented orthogonal basis may be used in the definition of ${\mbf \varepsilon}_{({\mathbb N})}$.

Projecting \eqref{pwsl6} onto $\mathbbm{1}$ and applying \eqref{epd11}$_2$ yields
\begin{equation}
\llbracket {\bf K}^{-1} \rrbracket \mathbbm{1} = (\mathbb{K}^\gamma)^{-1} - (\mathbb{K}^\delta)^{-1}= - {\mbf \beta}
_{r}^{T} {\mbf \varepsilon}_{({\mathbb N})} \label{pwsl6.1}
\end{equation}
which, on using ${\mbf \varepsilon}_{({\mathbb N})}^{2}=-\mathbbm{1}$, leads to
\begin{equation}
{\mbf \beta}_{r}^{T}\mathbbm{1}=\llbracket {\bf K}^{-1}\rrbracket {\mbf \varepsilon}_{({\mathbb N})}. \label{pwsl8}
\end{equation}
This determines the action of ${\mbf \beta}_{r}^{T}$ on $T_{S_r}({\bf X})$. The action of ${\mbf \beta}_{r}^{T}$ on $\mathbb N$ is indeterminate and may be set to zero without loss of generality. Consequently we assume ${\mbf \beta}_r^T$ to be superficial (i.e., ${\mbf \beta}_{r}^{T}\mathbbm{1} = {\mbf \beta}_{r}^{T}$) and write
\begin{equation}
{\mbf \beta}_{r}^T=  \llbracket {\bf K}^{-1}\rrbracket {\mbf \varepsilon}_{({\mathbb N})}. \label{pwsl9}
\end{equation}

The interface $S_r$ is coherent at ${\bf X} \in S_r$ if the Burgers vector as defined in \eqref{pwsl5} has no contribution from the line integral for all closed curves $C$ such that ${\bf X} \in \Gamma$. Accordingly, $S_r$ is coherent at ${\bf X} \in S_r$ if and only if the interface dislocation density ${\mbf \beta}_r$ at $\bf X$ vanishes.

Proceeding in parallel we can derive the spatial form of above relations. In particular
\begin{equation}
\llbracket {\bf H}^{-1} \rrbracket={\bf h} \otimes {\mathbbm n} -{\mbf \beta}_{t}^{T}{\mbf \varepsilon}_{({\mathbbm n})}~\text{and} ~{\mbf \beta}_{t}^{T}=\llbracket {\bf H}^{-1} \rrbracket {\mbf \varepsilon}_{({\mathbbm n})}, \label{pwsl12}
\end{equation}
where ${\bf h} \in \mathcal{V}$ is arbitrary and ${\mbf \varepsilon}_{({\mathbbm n})}$ is the two dimensional permutation tensor density on $T_{s_t}({\bf x})$. The tensor ${\mbf \beta}_{t}$ is
the (spatial) interface dislocation density which satisfies ${\mbf \beta}_{t}^{T} \bar{\mathbbm{1}} ={\mbf \beta}_{t}^{T}$. The Burgers vector using spatial description is given by
\begin{eqnarray}
{\bf b}(c,t) = \int_{c} {\bf H}^{-1}d{\bf x} &=& \int_{A_c}(\curl{\bf H}^{-1})^{T}{\bf n}_cda - \int_{\gamma
}\llbracket {\bf H}^{-1} \rrbracket d{\bf x} \nonumber
\\
&=& \int_{A_c}({\mbf \alpha}_t)^{T}{\bf n}_cda + \int_{\gamma}{\mbf \beta}_t^T \hat{\mathbbm t}_1 dl,
 \label{pwsl12.1}
\end{eqnarray}
where $c$ is a (closed) spatial curve enclosing area $A_c$ (with unit normal ${\bf n}_c$) and $\gamma = A_c \cap s_t$. The tensor ${\mbf \alpha}_t$ is to be identified with the (spatial) bulk dislocation density.  The curve $\gamma$ is parametrized by arc-length ${l}$ and has an associated tangent vector $\hat{\mathbbm t}_2$ such that the triad $\{\hat{\mathbbm t}_1,\hat{\mathbbm t}_2,{\mathbbm n}\}$ forms a positively oriented orthonormal basis on $s_t$.

The referential and spatial interface dislocation densities are not
independent. To obtain the relation start by noting that the jump in ${\bf F}^{-1}$ across the singular surfaces has the Hadamard's form $\llbracket {\bf F}^{-1}\rrbracket = {\bf a} \otimes {\mathbbm n}$ with ${\bf a} \in \mathcal{V}$ arbitrary. Using
this and \eqref{dai2} together with
\begin{equation}
\llbracket {\bf H}^{-1} \rrbracket= \langle {\bf K}^{-1} \rangle
\llbracket {\bf F}^{-1} \rrbracket + \llbracket {\bf K}^{-1} \rrbracket \langle {\bf F}^{-1} \rangle \label{pwsl13}
\end{equation}
we derive
\begin{equation}
{\bf h} \otimes {\mathbbm n} - {\mbf \beta}_{t}^{T}{\mbf \varepsilon}_{({\mathbbm n})}=\langle {\bf K}^{-1}\rangle
{\bf a} \otimes {\mathbbm n} + {\bf k} \otimes \langle {\bf F}^{-T} \rangle {\mathbb N} - {\mbf \beta}_{r}^{T} {\mbf \varepsilon}_{({\mathbb N})}\langle {\bf F}^{-1}\rangle . \label{pwsl14}
\end{equation}
Nanson's formula \eqref{sd17}$_1$ ensures that $\langle {\bf F}^{-T}\rangle {\mathbb N}$ is parallel to $\mathbbm n$. Multiplying \eqref{pwsl14} on the right by ${\mbf \varepsilon}_{({\mathbbm n})}$ then
furnishes ${\mbf \beta}_{t}^{T}$ in terms of ${\mbf \beta}_{r}^{T}$,
\begin{equation}
{\mbf \beta}_{t}^{T}=-{\mbf \beta}_{r}^{T} {\mbf \varepsilon}_{({\mathbb N})} \langle
{\bf F}^{-1} \rangle {\mbf \varepsilon}_{({\mathbbm n})}, \label{pwsl15}
\end{equation}
while the normal component of \eqref{pwsl14} yields a relationship among ${\bf a}$, ${\bf k}$, and $\bf h$:
\begin{equation}
{\bf h}=\langle {\bf K}^{-1} \rangle {\bf a} + ({\mathbbm n} \cdot \langle
{\bf F}^{-T} \rangle {\mathbb N}){\bf k}-{\mbf \beta}_{r}^{T} {\mbf \varepsilon}_{({\mathbb N})} \langle
{\bf F}^{-1} \rangle {\mathbbm n}. \label{pwsl16}
\end{equation}

\begin{rem} If ${\bf K}^{-1}$ is the gradient of a piecewise continuously differentiable deformation ${\mbf \chi}^p$, i.e. ${\bf K}^{-1} = \nabla {\mbf \chi}^p$, then \cite{TruesdellToupin60}
\begin{equation}
\llbracket {\bf K}^{-1}\rrbracket = {\bf k} \otimes {\mathbb N} + \nabla^S \llbracket {\mbf \chi}^p \rrbracket. \label{pwsl10}
\end{equation}
The interface dislocation density, given by \eqref{pwsl8}, then has the form
\begin{equation}
{\mbf \beta}_{r}^{T}\mathbbm{1}=\nabla^S \llbracket {\mbf \chi}^p\rrbracket {\mbf \varepsilon}_{({\mathbb N})} = \nabla \llbracket {\mbf \chi}^p\rrbracket {\mbf \varepsilon}_{({\mathbb N})}, \label{pwsl11}
\end{equation}
where we have used the identity $\mathbbm{1} {\mbf \varepsilon}_{({\mathbb N})} = {\mbf \varepsilon}_{({\mathbb N})}$. This situation occurs when the bulk dislocation density ${\mbf \alpha}^r = \Curl{\bf K}^{-1}$ vanishes on either side of the interface and the dislocation distribution is restricted to the singular surface $S_r$. An equivalent formulation holds with respect to the spatial configuration. The previous work on incoherent interfaces \cite{CermelliGurtin94, LeoSekerka89} has been in fact restricted to this case with the exception of the paper by Cermelli and Sellers \cite{CermelliSellers00} who have developed the theory in the context of crystal lattice vectors.
\end{rem}

\begin{rem}(Interface dislocation nodes) An interface dislocation node, as introduced by Bilby \cite{Bilby55}, is the line of intersection of interfaces with dislocation density distributions. The analysis in \cite{Bilby55} is restricted to plane interfaces and infinitesimal strains. We extend it for curved surfaces and finite distortions. Consider $N$ surfaces intersecting at a line $L \subset \kappa_r$. Each surface has an associated normal and a distribution of interface dislocation density. The following compatibility relation holds in a neighborhood infinitesimal close to $L$
\begin{equation}
\overset{N}{\underset{i=1}{\sum}} \llbracket {\bf K}^{-1 (i)} \rrbracket = {\bf 0}, \label{pwsl17}
\end{equation}
where the index $i$ in the superscript represents the $i$'th interface. This relation follows from the observation that on passing around the line $L$ (in a small neighborhood) one reaches the initial material point; and $({\bf K}^{-1 (i)})^+ = ({\bf K}^{-1 (i+1)})^-$, where $(i+1)$ should be taken as $1$ when $i=N$.

Use \eqref{pwsl6} to rewrite \eqref{pwsl17} as
\begin{equation}
\overset{N}{\underset{i=1}{\sum}}( {\bf k}^{(i)} \otimes {\mathbb N}^{(i)} - {\mbf \beta}
_{r}^{T(i)} {\mbf \varepsilon}_{({\mathbb N}^{(i)})}) = {\bf 0}. \label{pwsl18}
\end{equation}
Let ${\mathbbm t}$ be the unit tangent vector field associated with line $L$. We can therefore choose vector ${\mathbbm t}^{(i)} \in T_{S_r^{(i)}}$ at ${\bf X} \in L$ such that $\{{\mathbbm t},{\mathbbm t}^{(i)},{\mathbb N}^{(i)}\}$ forms a positively oriented orthonormal basis at ${\bf X}\in L$ for each intersecting surface. We also have, cf. \eqref{pwsl7},
\begin{equation}
{\mbf \varepsilon}_{({\mathbb N}^{(i)})} = {\mathbbm t} \otimes {\mathbbm t}^{(i)} - {\mathbbm t}^{(i)} \otimes {\mathbbm t}. \label{pwsl19}
\end{equation}
On substituting this in \eqref{pwsl18} it follows immediately that
\begin{equation}
{\bf 0} = \overset{N}{\underset{i=1}{\sum}} \llbracket {\bf K}^{-1 (i)} \rrbracket {\mathbbm t} = \overset{N}{\underset{i=1}{\sum}}{\mbf \beta}
_{r}^{T(i)} {\mathbbm t}^{(i)} \label{pwsl20}
\end{equation}
at ${\bf X}\in L$. The outer equality in \eqref{pwsl20} provides us with a compatibility condition relating the interface dislocation density tensors of various intersecting surfaces. This can be compared to the equation of conservation of Burgers vectors. Equivalently, in terms of the spatial surface dislocation density, we can obtain $\overset{N}{\underset{i=1}{\sum}}{\mbf \beta}
_{t}^{T(i)} \hat{\mathbbm t}^{(i)} = {\bf 0}$, where $\hat{\mathbbm t}^{(i)} \in T_{s_t^{(i)}}$.

\label{surnodes}
\end{rem}

\subsection{\label{trsurdd}True interface dislocation density}

A measure of the bulk dislocation density, invariant with respect to compatible changes in the reference configuration, is given by \cite{CermelliGurtin01} \begin{equation}
J_{K} {\bf K}^{-1}\Curl {\bf K}^{-1}={\mbf \alpha}=J_{H} {\bf H}^{-1} \curl {\bf H}^{-1}. \label{dai6}
\end{equation}
Tensor $\mbf \alpha$ is a map from a local configuration in $\mathcal{M} \setminus \mathcal{N}$ onto itself. It is, therefore, also invariant under superimposed rigid body motions. The importance of $\mbf \alpha$ is perhaps most evident in its appearance in constitutive response functions related to plastic flow (cf. \cite{EpsteinElz07, Guptaetal07} and \eqref{fay6} below). Response functions, if invariant under compatible changes in the reference configuration, cannot depend explicitly on $\bf K$ and their dependence on $\nabla{\bf K}$ is only through $\mbf \alpha$ \cite{CermelliGurtin01, Davini88}.

We now obtain an analogous measure for interface dislocation density. Consider two reference configurations related by a compatible deformation with smooth gradient $\bf A$ (refer to Subsection \ref{ssm} for the kinematics of a compatible change in the reference configuration). We impose the requirement that under a compatible transformation the Burgers vector is left invariant and consequently
\begin{equation}
\int_{A_{C_1}}(\Curl_1~{\bf K}_1^{-1})^{T}{\bf N}_{C_1} dA_1 - \int_{\Gamma_1
}\llbracket {\bf K}_1^{-1} \rrbracket d{\bf X}_1 = \int_{A_{C_2}}(\Curl_2~{\bf K}_2^{-1})^{T}{\bf N}_{C_2} dA_2 - \int_{\Gamma_2
}\llbracket {\bf K}_2^{-1} \rrbracket d{\bf X}_2, \label{pwsl21}
\end{equation}
where $A_{C_2} = {\mbf \lambda}(A_{C_1})$, $\Gamma_2 = {\mbf \lambda}(\Gamma_1)$, and ${\bf K}_2 = {\bf A}{\bf K}_1$ (cf. \eqref{invcomp1}). The area integrals in \eqref{pwsl21} are equal, owing to the invariance of $\mbf \alpha$, thereby reducing \eqref{pwsl21} to
\begin{equation}
\int_{\Gamma_1
}\llbracket {\bf K}_1^{-1} \rrbracket d{\bf X}_1 = \int_{\Gamma_2
}\llbracket {\bf K}_2^{-1} \rrbracket d{\bf X}_2. \label{pwsl22}
\end{equation}
Let ${\mathbbm t}_a$ and $L_a$ be the unit tangent vector and the arc-length associated with $\Gamma_a$, respectively ($a=1,2$). Then
\begin{equation}
{\mathbbm t}_2 dL_2 = d{\bf X}_2 = {\bf A} d{\bf X}_1 = {\bf A} {\mathbbm t}_1 dL_1.  \label{pwsl23}
\end{equation}
Substitute $d{\bf X}_2$ from \eqref{pwsl23} into \eqref{pwsl22} and employ the arbitrariness of $\Gamma_1$ to obtain
\begin{equation}
\llbracket {\bf K}_2^{-1} \rrbracket {\bf A} {\mathbbm t}_1 = \llbracket {\bf K}_1^{-1}\rrbracket {\mathbbm t}_1 \label{pwsl24}
\end{equation}
for ${\mathbbm t}_1 \in T_{S_{r_1}}$. Because ${\mathbbm t}_1$ is otherwise arbitrary this relation is satisfied only if
\begin{equation}
\llbracket {\bf K}_2^{-1} \rrbracket {\bf A} - \llbracket {\bf K}_1^{-1}\rrbracket = {\bf c} \otimes {\mathbb N}_1, \label{pwsl25}
\end{equation}
where ${\bf c}\in \mathcal{V}$ is arbitrary. Project both sides of \eqref{pwsl25} on $\mathbbm{1}_1$  and use \eqref{pwsl6.1} to get
\begin{equation}
{\mbf \beta}_{r_2}^T {\mbf \varepsilon}_{({\mathbb N}_2)} {\mathbb A}  = {\mbf \beta}_{r_1}^T {\mbf \varepsilon}_{({\mathbb N}_1)} \label{pwsl26}
\end{equation}
or equivalently, on applying \eqref{invcomp1.02}$_1$,
\begin{equation}
{\mbf \beta}_{r_2}^T {\mbf \varepsilon}_{({\mathbb N}_2)} {\mathbb K}^\alpha_2  = {\mbf \beta}_{r_1}^T {\mbf \varepsilon}_{({\mathbb N}_1)} {\mathbb K}^\alpha_1 \label{pwsl26.1}
\end{equation}
with $\alpha = \{{\gamma, \delta}\}$.
This leads us to define invariant (or true) interface dislocation densities ${\mbf \beta}^\gamma$ and ${\mbf \beta}^\delta$ by
\begin{equation}
{\mbf \beta}^{\gamma} = {\mbf \beta}_{r}^T {\mbf \varepsilon}_{({\mathbb N})} {\mathbb K}^\gamma~\text{and}~{\mbf \beta}^{\delta} = {\mbf \beta}_{r}^T {\mbf \varepsilon}_{({\mathbb N})} {\mathbb K}^\delta. \label{pwsl38}
\end{equation}

Define the relative distortion tensor
\begin{equation}
{\mathbb M} = ({\mathbb H}^\gamma)^{-1} {\mathbb H}^\delta = ({\mathbb K}^\gamma)^{-1} {\mathbb K}^\delta \label{ied4}
\end{equation}
which is a linear map between two local configurations at a material point on the incoherent interface. It was introduced by Ceremelli and Gurtin \cite{CermelliGurtin94, CermelliGurtin94b} where it was called the incoherency tensor. The second equality in \eqref{ied4} follows from \eqref{epd8}. It is checked easily that $\mathbb{M}$ satisfies $\mathbb{M} = \mathbb{M} \mathbbm{1}^\delta$ and $\mathbb{M} = \mathbbm{1}^\gamma \mathbb{M}$. Also, observe that
\begin{equation}
{\mbf \beta}^{\delta} = {\mbf \beta}^{\gamma} \mathbb{M}. \label{pwsl38.2}
\end{equation}
Substitute ${\mbf \beta}_{r}^T {\mbf \varepsilon}_{({\mathbb N})}$ from \eqref{pwsl6.1}$_2$ and use \eqref{ied4} to get
\begin{equation}
{\mbf \beta}^{\gamma} =\mathbb{M}^{-1} - {\mathbbm 1}^\gamma~\text{and}~{\mbf \beta}^{\delta} = {\mathbbm 1}^\delta - \mathbb{M}, \label{pwsl38.3}
\end{equation}
where $\mathbb{M}^{-1}$ is the pseudoinverse of $\mathbb{M}$ such that $\mathbb{M}^{-1}\mathbb{M}= \mathbbm{1}^\delta$ and $\mathbb{M}\mathbb{M}^{-1}= \mathbbm{1}^\gamma$.

If in the above analysis we substitute $\bf F$ for $\bf A$ we obtain, instead of \eqref{pwsl26.1}, the following relation
\begin{equation}
{\mbf \beta}_{t}^T {\mbf \varepsilon}_{({\mathbbm n})} {\mathbb H}^\alpha  = {\mbf \beta}_{r}^T {\mbf \varepsilon}_{({\mathbb N})} {\mathbb K}^\alpha \label{pwsl38.1}
\end{equation}
with $\alpha = \{{\gamma, \delta}\}$. The true interface dislocation densities can thus be equivalently expressed in terms of elastic distortion.

For the incoherent interface the tangent plane (to the singular surface) in the reference (or spatial) configuration is mapped (locally) into two
tangent planes in the relaxed manifold. As a result we
have two measures, ${\mbf \beta}^\gamma$ and ${\mbf \beta}^\delta$, of true interface dislocation density for each ${\bf
X}\in T_{S_r}$ (or ${\bf x}\in T_{s_t}$). This is in contrast to the bulk where, as pointed out in the beginning of this Subsection, we have a single measure of invariant bulk dislocation density.

\begin{rem} (Equivalence between different dislocation densities) With different measures of dislocation density distributions, both in bulk and on interface, it is useful to investigate the possibility of equivalence between them. The true bulk dislocation density, $\mbf \alpha$, vanishes if and only if ${\mbf \alpha}_r$ (or ${\mbf \alpha}_t$) vanishes. This is evident from \eqref{dai6}, where the determinants are positive and distortions invertible. This however is not the case with their time derivatives. Similarly, the true interface dislocation densities vanish if and only if ${\mbf \beta}_r$ (or ${\mbf \beta}_t$) vanishes. This follows on using \eqref{epd9} and \eqref{epd10} in \eqref{pwsl38} and \eqref{pwsl38.1}. Their normal time derivative however might not all be zero at the same instant.
\end{rem}

\subsection{\label{jbdd}Relationship between bulk and interface dislocation densities}

We now relate the jump in bulk dislocation density across the interface with interface dislocation density. An integral form of conservation of dislocations, for an arbitrary volume which intersects the singular interface, is also obtained. Our discussion involves referential dislocation densities; similar relations can be obtained using their
spatial counterparts.

A compatibility condition for  $\nabla {\bf K}^{-1}$, discontinuous across $S_r$ but otherwise smooth, is given by \cite{Thomas61, TruesdellToupin60}
\begin{equation}
\llbracket \nabla {\bf K}^{-1} \rrbracket = {\bf Q} \otimes {\mathbb N} + \nabla^S \llbracket {\bf K}^{-1} \rrbracket, \label{pwsl39}
\end{equation}
where ${\bf Q} \in Lin$ is arbitrary. In terms of indicial notation this is alternatively written as
\begin{equation}
\llbracket K^{-1}_{jl,k} \rrbracket = Q_{jl} {\mathbb N}_k + \llbracket K^{-1}_{jl} \rrbracket_{,m} \mathbbm{1}_{mk}. \label{pwsl40}
\end{equation}
Multiply \eqref{pwsl40} throughout by $e_{ikl}$ and use the definition of referential bulk dislocation density to obtain (the subscript present, if any, in the bold notation is written as a superscript in the indicial notation)
\begin{equation}
\llbracket \alpha^r_{ij} \rrbracket = e_{ikl} Q_{jl} {\mathbb N}_k + e_{ikl} \llbracket K^{-1}_{jl} \rrbracket_{,m} \mathbbm{1}_{mk}. \label{pwsl41}
\end{equation}
Using $e_{ikl} {\mathbb N}_i {\mathbb N}_k = 0$, the normal projection of \eqref{pwsl41} yields
\begin{equation}
\llbracket \alpha^r_{ij} \rrbracket {\mathbb N}_i =  e_{ikl} \llbracket K^{-1}_{jl} \rrbracket_{,k} {\mathbb N}_i. \label{pwsl45}
\end{equation}
The jump in ${\bf K}^{-1}$, given in \eqref{pwsl6}, can be written in indicial notation as
\begin{equation}
\llbracket K^{-1}_{ij} \rrbracket = {k}_i {\mathbb N}_j + \beta^r_{ki} \epsilon^{({\mathbb N})}_{kj}. \label{pwsl46}
\end{equation}
Replace $\llbracket K^{-1}_{ij} \rrbracket$ from \eqref{pwsl46} in \eqref{pwsl45} to get
\begin{equation}
\llbracket \alpha^r_{ij} \rrbracket {\mathbb N}_i =  e_{ikl} k_j {\mathbb N}_{l,k} {\mathbb N}_i + e_{ikl} (\beta^r_{qj} \epsilon^{({\mathbb N})}_{ql} )_{,k} {\mathbb N}_i. \label{pwsl47}
\end{equation}
The first term on the right hand side vanishes since $e_{ikl} {\mathbb N}_{l,k} {\mathbb N}_i =  0$, which can be proved using \eqref{s2}$_1$ and the skew symmetry of $e_{ikl}$. Consequently \eqref{pwsl47} reduces to
\begin{equation}
\llbracket \alpha^r_{ij} \rrbracket {\mathbb N}_i =  \beta^r_{qj,k} e_{ikl} \epsilon^{({\mathbb N})}_{ql} {\mathbb N}_i + \beta^r_{qj} e_{ikl} \epsilon^{({\mathbb N})}_{ql,k} {\mathbb N}_i. \label{pwsl49}
\end{equation}
We note the following two identities:
\begin{eqnarray}
&& e_{ikl} \epsilon^{({\mathbb N})}_{ql} {\mathbb N}_i = \mathbbm{1}_{qk}~\text{and} \label{pwsl49.1}
\\
&& e_{ikl} \epsilon^{({\mathbb N})}_{ql,k} {\mathbb N}_i = 2H {\mathbb N}_q. \label{pwsl49.2}
\end{eqnarray}
Relation \eqref{pwsl49.1} follows from the definition of $\epsilon^{({\mathbb N})}_{ql}$. The proof for \eqref{pwsl49.2}, which is left to the reader, is however more involved and requires calculating divergence of \eqref{pwsl49.1} and using $\nabla{\mathbb N} = -\mathbb{L} + (\nabla{\mathbb N}){\mathbb N}\otimes {\mathbb N}$, where $\mathbb{L}$ is the symmetric curvature tensor defined in \eqref{sd3}.

Use \eqref{pwsl49.1} and \eqref{pwsl49.2} to write \eqref{pwsl49} equivalently as $\llbracket {\mbf \alpha}^T_r \rrbracket {\mathbb N} =  \Div^S {\mbf \beta}^T_r +  2H {\mbf \beta}^T_r {\mathbb N}$ or
\begin{equation}
\llbracket {\mbf \alpha}^T_r \rrbracket {\mathbb N} =  \Div^S {\mbf \beta}^T_r ~\forall {\bf X} \in S_r, \label{pwsl49.9}
\end{equation}
given that ${\mbf \beta}^T_r$ is superficial. This is an important result highlighting the nature of interface dislocation densities (cf. $\Div^S {\mbf \alpha}^T_r = {\bf 0}~ \forall {\bf X} \in \kappa_r \setminus S_r$). To expand on this we use the surface divergence theorem \eqref{st6.1}. For an arbitrary surface $S \subset S_r$ we have
\begin{equation}
\int_{\partial  S} {\mbf \beta}^T_r {\mbf \nu} dL = \int_{S} \Div^S {\mbf \beta}^T_r dA, \label{pwsl50.1}
\end{equation}
where $\mbf \nu$ is the outer unit normal to ${\partial S}$. In addition, use the divergence theorem for piecewise smooth fields, and the identity $\Div{\mbf \alpha}^T_r = {\bf 0}$ away from $S$, to derive
\begin{equation}
\int_{\partial  \Omega} {\mbf \alpha}^T_r {\bf N} dA = \int_{S} \llbracket {\mbf \alpha}^T_r \rrbracket {\mathbb N}  dA, \label{pwsl50.4}
\end{equation}
where $\bf N$ is the normal to $\partial \Omega$. Combining \eqref{pwsl49.9}, \eqref{pwsl50.1}, and \eqref{pwsl50.4} yields
\begin{equation}
\int_{\partial  S} {\mbf \beta}^T_r {\mbf \nu} dL = \int_{\partial  \Omega} {\mbf \alpha}^T_r {\bf N} dA \label{pwsl50.5}
\end{equation}
as the integral law for conservation of dislocations in an arbitrary volume $\Omega \subset \kappa_r$ such that $S = S_r \cap \Omega \neq \emptyset$.

If $\llbracket {\mbf \alpha}^T_r \rrbracket {\mathbb N} = {\bf 0}$, i.e. there are no external sources to interfacial dislocation density, then \eqref{pwsl50.5} reduces to
\begin{equation}
\int_{\partial  S} {\mbf \beta}^T_r {\mbf \nu} dL = {\bf 0}, \label{pwsl50.2}
\end{equation}
which can be interpreted as the conservation law for interface dislocations. This is analogous to the conservation law for ${\mbf \alpha}_r$ according to which, for an arbitrary volume $\Omega \subset \kappa_r$ with $S_r \cap \Omega = \emptyset$,
\begin{equation}
\int_{\partial  \Omega} {\mbf \alpha}^T_r {\bf N} dA = {\bf 0}. \label{pwsl50.3}
\end{equation}
Relation \eqref{pwsl50.3} imposes the restriction on bulk dislocations to not end arbitrarily inside $\Omega$. A parallel interpretation in the context of interface dislocation densities is furnished by  \eqref{pwsl50.2}. Therefore for a vanishing normal jump in ${\mbf \alpha}_r^T$ an incoherent interface cannot end arbitrarily inside the solid. Such an interface will either end at the boundary of the solid or at a surface dislocation
node (see Remark \ref{surnodes} above).

\begin{rem} For moving interfaces ($U \neq 0$) we can derive an alternate expression for the jump in bulk dislocation density.
The jump in $\dot{K^{-1}_{jl}}$ can be expressed in terms of the normal time derivative of $\llbracket K^{-1}_{jl} \rrbracket$ as
\begin{equation}
\llbracket K^{-1}_{jl}~ \rrbracket{\mathring{}} =  U\llbracket K^{-1}_{jl,k} \rrbracket {{\mathbb N}_k} + \llbracket \dot{K^{-1}_{jl}} \rrbracket,  \label{pwsl42}
\end{equation}
which is obtained by subtracting the normal time derivative of $ \left(K^{-1}_{jl}\right)^-$ from that of $ \left(K^{-1}_{jl}\right)^+$. Substituting \eqref{pwsl40} into \eqref{pwsl42} yields
\begin{equation}
U Q_{jl}  = \llbracket K^{-1}_{jl}~ \rrbracket{\mathring{}} - \llbracket \dot{K^{-1}_{jl}} \rrbracket.  \label{pwsl43}
\end{equation}
Replace $\bf Q$ in \eqref{pwsl41} from \eqref{pwsl43} to obtain
\begin{equation}
U \llbracket \alpha^r_{ij} \rrbracket = - e_{ikl}\llbracket \dot{K^{-1}_{jl}} \rrbracket {\mathbb N}_k + e_{ikl}\llbracket K^{-1}_{jl} ~\rrbracket{\mathring{}}~ {\mathbb N}_k + U e_{ikl} \llbracket K^{-1}_{jl} \rrbracket_{,m} \mathbbm{1}_{mk} \label{pwsl44}
\end{equation}
as the jump condition for bulk dislocation density across a moving interface. This relation reveals various sources in the production of bulk dislocation density (near the interface) as the interface moves with velocity $U$. Observe that $\llbracket {\mbf \alpha}_r \rrbracket$ does not necessarily vanish for continuous ${\bf K}^{-1}$.
\end{rem}

\section{\label{ipdkl}Interfacial plasticity: dissipation, energetics, and kinetics}

Plastic flow is a dissipative process involving irreversible restructuring of the microstructure which in turn affects the macroscopic behavior of bodies. Away from the interface the dissipation is caused by evolution of plastic distortion. At the interface there are three dissipative mechanisms: the motion of interface (governed by its normal velocity), the evolution of plastic distortion, and the evolution of relative plastic distortion (incoherency). All these will, in general, be coupled to each other. In the following we start with specific constitutive assumptions on the nature of energy densities and stresses and then use them to evaluate the dissipation. The driving forces for various dissipative mechanisms are obtained. Based on the list of dissipative fluxes and driving forces general forms for kinetic laws are proposed and then simplified using invariance requirements under various symmetries.

\subsection{\label{pidi}Dissipation inequality}

We now revisit dissipation inequalities \eqref{di1} and \eqref{di2} assuming energy densities such that \cite{Guptaetal07}
\begin{eqnarray}
&& \Psi = J_K^{-1} W({\bf H})~\text{and}  \label{tdi1}
\\
&& \Phi = ({j^\gamma})^{-1} w({\mathbb H}^\gamma, {\mathbb H}^\delta), \label{tdi2}
\end{eqnarray}
where $W$ is the bulk energy density per unit volume of a relaxed configuration $\kappa_i$ and $w$ is
the interface energy per unit area of a local configuration $T_{\mathcal N}^\gamma$ (we can equivalently consider an interfacial energy per unit area of $T_{\mathcal N}^\delta$). While it is possible to include higher-order gradients at the interface to reflect bending and other weakly non-local
effects (as in \cite{SteigmannOgden99}), it is our view that the present 'membrane-like' model, in the spirit of the
Gurtin-Murdoch model of interfaces \cite{GurtinMurdoch75}, represents the leading-order effects
faithfully if the actual interfacial region is sufficiently thin (as distinct from our present representation as a
discontinuity surface). This issue represents a direction for future research in interfacial plasticity.

Under the hypothesis of hyperelastic response (during elastic unloading), the bulk and interfacial Cauchy stresses are assumed to be
\begin{eqnarray}
&& J_H {\bf T}= W_{\bf H} {\bf H}^T~\text{and} \label{tdi3}
\\
&& j_s^\gamma {\mathbb T} = \sum_{\alpha=\gamma,\delta} w_{\mathbb{H}^\alpha} \left({\mathbb{H}^\alpha}\right)^T, \label{tdi4}
\end{eqnarray}
respectively, where the summation is over both the local configurations in $\mathcal{N}$ at a fixed material point on the interface. Recalling ${\bf P} = {\bf T}{\bf F}^\ast$,
in addition to \eqref{bl26.3}, \eqref{epd6.1}, and \eqref{epd8}, leads to the corresponding Piola stresses
\begin{eqnarray}
&& J_K {\bf P}= W_{\bf H} {\bf K}^T~\text{and} \label{tdi5}
\\
&& j^\gamma {\mathbb P} = \sum_{\alpha=\gamma,\delta} w_{\mathbb{H}^\alpha} \left({\mathbb{K}^\alpha}\right)^T. \label{tdi6}
\end{eqnarray}
We note that relations \eqref{tdi4} and \eqref{tdi6} are motivated by the assumption that the total interfacial Piola stress ${\mathbb P}$ is power-conjugate to interfacial deformation gradient ${\mathbb F}$, as in \eqref{tdi12} below.

The dissipation inequality in the bulk \eqref{di1}, on using \eqref{tdi1} and \eqref{tdi3}, reduces to (see for example \cite{EpsteinMaugin90, Guptaetal07})
\begin{equation}
{\bf E} \cdot \dot{\bf K}{\bf K}^{-1}  \geq 0~\forall {\bf X} \in \kappa_r \setminus S_r, \label{tdi6.1}
\end{equation}
where ${\bf E}$ is the bulk Eshelby tensor defined in \eqref{cdi23}.

We now evaluate interface dissipation inequality \eqref{di2} under the above constitutive assumptions. Use \eqref{tdi2} to obtain the normal time derivative of interface energy density
\begin{equation}
\mathring{\Phi} = -\mathring{j}^\gamma \left({j}^\gamma\right)^{-1} \Phi + \left({j}^\gamma\right)^{-1} \sum_{\alpha=\gamma,\delta} w_{\mathbb{H}^\alpha} \cdot \mathring{\mathbb{H}}^\alpha . \label{tdi7}
\end{equation}
Additionally, note that $w_{\mathbb{H}^\alpha} {\mathbb N}^\alpha = {\bf 0}$ (no summation) for ${\alpha=\{\gamma,\delta}\}$ (cf. \eqref{deronsur2}$_1$).  Substituting $\mathring{j}^\gamma$ from \eqref{epd12}$_3$ into \eqref{tdi7} then yields
\begin{equation}
\mathring{\Phi} = - \mathring{\mathbb{K}}^\gamma({\mathbb{K}^\gamma})^{-1} \cdot {\mathbbm 1}  \Phi + \left({j}^\gamma\right)^{-1} \sum_{\alpha=\gamma,\delta} w_{\mathbb{H}^\alpha} \mathbbm{1}^\alpha \cdot \mathring{\mathbb{H}}^\alpha. \label{tdi8}
\end{equation}
Taking the normal time derivative of \eqref{epd8} and using it to replace $ \mathring{\mathbb{H}^\alpha}$ above leads to
\begin{equation}
 \mathring{\Phi}= - \mathring{\mathbb{K}}^\gamma({\mathbb{K}^\gamma})^{-1} \cdot {\mathbbm 1}  \Phi + \left({j}^\gamma\right)^{-1} \sum_{\alpha=\gamma,\delta} \left( w_{\mathbb{H}^\alpha} ({\mathbb{K}^\alpha})^T \cdot \mathring{\mathbb{F}} + \mathbb{F}^T w_{\mathbb{H}^\alpha} {\mathbbm{1}^\alpha} \cdot \mathring{\mathbb{K}}^\alpha\right). \label{tdi9}
\end{equation}
or equivalently (recall \eqref{tdi6} and \eqref{epd10}$_1$)
\begin{equation}
\mathring{\Phi} = - \mathring{\mathbb{K}}^\gamma({\mathbb{K}^\gamma})^{-1} \cdot {\mathbbm 1}  \Phi + {\mathbb P} \cdot \mathring{\mathbb{F}} + \left({j}^\gamma\right)^{-1} \sum_{\alpha=\gamma,\delta} \left( \mathbb{F}^T w_{\mathbb{H}^\alpha} ({\mathbb{K}^\alpha})^T \cdot \mathring{\mathbb{K}}^\alpha({\mathbb{K}^\alpha})^{-1}\right). \label{tdi10}
\end{equation}
This can be expressed succinctly on introducing
\begin{equation}
j^\gamma \mathbb{P}^\gamma = w_{\mathbb{H}^\gamma} ({\mathbb{K}^\gamma})^T~\text{and}~j^\gamma \mathbb{P}^\delta = w_{\mathbb{H}^\delta} ({\mathbb{K}^\delta})^T \label{tdi11}
\end{equation}
as two components of the interface Piola stress such that $\mathbb{P} = \mathbb{P}^\gamma + \mathbb{P}^\delta$. Equation \eqref{tdi10} can be then rewritten as
\begin{equation}
\mathring{\Phi} =  {\mathbb P} \cdot \mathring{\mathbb{F}} - \mathbb{E}^\gamma \cdot \mathring{\mathbb{K}}^\gamma({\mathbb{K}^\gamma})^{-1} + \mathbb{F}^T  \mathbb{P}^\delta \cdot \mathring{\mathbb{K}}^\delta({\mathbb{K}^\delta})^{-1}, \label{tdi12}
\end{equation}
where
\begin{equation}
\mathbb{E}^\gamma = \Phi \mathbbm{1} - \mathbb{F}^T  \mathbb{P}^\gamma \label{tdi13}
\end{equation}
is the interface Eshelby tensor associated with $T_\mathcal{N}^\gamma$. Define $\mathbb{E}^\delta$ such that
\begin{equation}
\mathbb{E} = \mathbb{E}^\gamma + \mathbb{E}^\delta, \label{tdi14}
\end{equation}
where $\mathbb{E}$ is the total interface Eshelby tensor introduced in \eqref{cdi23.1}. Hence
\begin{equation}
\mathbb{E}^\delta =  - \mathbb{F}^T  \mathbb{P}^\delta. \label{tdi15}
\end{equation}
The apparent asymmetry in the definition of two interface Eshelby tensors, in \eqref{tdi13} and \eqref{tdi15}, is due to our use of an interface energy density measured per unit area of $T_\mathcal{N}^\gamma$. The dissipation inequality \eqref{di2}, on replacing $\mathring{\Phi}$ from \eqref{tdi12}, acquires the form
\begin{equation}
U \left({\mathbb N} \cdot \llbracket {\bf E} \rrbracket {\mathbb N} +
\frac{1}{2} U^2 \rho_r \llbracket | {\bf F}{\mathbb N} |^2 \rrbracket
 +  \mathbb{E} \cdot \mathbb{L} \right) +  \sum_{\alpha=\gamma,\delta} \mathbb{E}^\alpha \cdot \mathring{\mathbb{K}}^\alpha({\mathbb{K}^\alpha})^{-1} \geq 0 ~  \forall {\bf
X} \in S_r, \label{tdi16}
\end{equation}
or
\begin{equation}
U \mathbbm{f} +  \sum_{\alpha=\gamma,\delta} \mathbb{E}^\alpha \cdot \mathring{\mathbb{K}}^\alpha({\mathbb{K}^\alpha})^{-1} \geq 0 ~  \forall {\bf
X} \in S_r, \label{tdi17}
\end{equation}
where
\begin{equation}
\mathbbm{f} = \left({\mathbb N} \cdot \llbracket {\bf E} \rrbracket {\mathbb N} +
\frac{1}{2} U^2 \rho_r \llbracket | {\bf F}{\mathbb N} |^2 \rrbracket
\right) +  \mathbb{E} \cdot \mathbb{L} \label{tdi18}
\end{equation}
is the driving force associated with the normal motion of the interface and $\mathbb{L}$ is the curvature tensor. The left side of inequality \eqref{tdi17} is the net dissipation caused by a moving interface and interfacial plastic flow. For a coherent interface the two local configurations $T_\mathcal{N}^\gamma$ and $T_\mathcal{N}^\delta$ coincide, and ${\mathbb{K}}^\gamma = {\mathbb{K}}^\delta$ ($={\mathbb{K}}$, say). The dissipation inequality \eqref{tdi17} then takes the form
\begin{equation}
U \mathbbm{f} + \mathbb{E} \cdot \mathring{\mathbb{K}}({\mathbb{K}})^{-1} \geq 0 ~  \forall {\bf
X} \in S_r, \label{tdi17c}
\end{equation}

The interface Eshelby tensor appears naturally as the driving force for plastic flow at the interface. This is comparable to the plastic behavior away from the interface, cf. \eqref{tdi6.1}, where the bulk Eshelby tensor drives the bulk plastic flow. Moreover, it is clear from \eqref{tdi17} that the normal projections of plastic distortion rates, i.e. $\mathring{\mathbb{K}}^\alpha \mathbb{N}^\alpha$, do not participate in dissipation. This is not the case when the energy density depends explicitly on interface normals, as discussed briefly in Remark \ref{normdep} below.

\begin{rem}(Area and volume preserving plastic flow) Assume $J_K = 1$ and $j^\alpha = 1$ for $\alpha = \{\gamma,\delta\}$. Therefore, the plastic distortion brings about no change in the volume and the area of bulk and interface, respectively. The dissipation inequalities \eqref{tdi6.1} and \eqref{tdi17} are reduced to
\begin{eqnarray}
&& {\bf F}^T {\bf P} \cdot \dot{\bf K}{\bf K}^{-1}  \leq 0~\forall {\bf X} \in \kappa_r \setminus S_r~\text{and} \label{tdi19}
\\
&& U \mathbbm{f} - \sum_{\alpha=\gamma,\delta} \mathbb{F}^T  \mathbb{P}^\alpha \cdot \mathring{\mathbb{K}}^\alpha({\mathbb{K}^\alpha})^{-1} \geq 0 ~~  \forall {\bf
X} \in S_r. \label{tdi20}
\end{eqnarray}
This form is similar to classical plasticity theories wherein the plastic flow is driven by stress rather than the Eshelby tensor.
\end{rem}

\begin{rem} Consider an interfacial energy density of the form $\check{w}({\mathbb H}^\gamma, {\mathbb H}^\delta, {\mathbb N}^\gamma, \mathbb{N}^\delta)$ with explicit dependence on the interfacial normals. Such energies (for coherent and unstrained interface) have been used for example to model the anisotropy of the surface during crystal growth \cite{Herring51}. The normal projections $\check{w}_{{\mathbb H}^\alpha} \mathbb{N}^\alpha$ (for each $\alpha = \{\gamma,\delta\}$) are no longer zero, cf. \eqref{deronsur2}$_1$, and they contribute to the driving force for the normal evolution of plastic distortion. Indeed, the term
\begin{equation}
\sum_{\alpha=\gamma,\delta} \left( \check{w}_{{\mathbb H}^\alpha} \mathbb{N}^\alpha \cdot \mathring{\mathbb H}^\alpha \mathbb{N}^\alpha + \check{w}_{{\mathbb N}^\alpha} \cdot \mathring{\mathbb N}^\alpha \right) \label{tdi21}
\end{equation}
has to be now appended to the left hand side of the inequality \eqref{tdi17}. Take normal time derivatives of ${\mathbb H}^\alpha {\mathbb N}^\alpha = {\bf 0}$ and ${\mathbb K}^\alpha {\mathbb N}^\alpha = {\bf 0}$ to obtain  $\mathring{\mathbb H}^\alpha \mathbb{N}^\alpha = -{\mathbb H}^\alpha \mathring{\mathbb{N}}^\alpha$ and $\mathring{\mathbb{N}}^\alpha = -({\mathbb K}^\alpha)^{-1} \mathring{\mathbb K}^\alpha \mathbb{N}^\alpha$, respectively. Using these we can rewrite \eqref{tdi21} as
\begin{equation}
\sum_{\alpha=\gamma,\delta} {\mathbbm c}^\alpha \cdot (\mathbb{K}^\alpha)^{-1} \mathring{\mathbb K}^\alpha \mathbb{N}^\alpha,  \label{tdi22}
\end{equation}
where $\mathbbm{c}^\alpha = (\mathbb{H}^\alpha)^T \check{w}_{\mathbb{H}^\alpha} {\mathbb N}^\alpha -  \check{w}_{\mathbb{N}^\alpha}$, cf. paragraph before \eqref{di2.002}.
\label{normdep}
\end{rem}

\paragraph{Interface energy density} Invariance of $w$ under a superimposed rigid body rotation requires
\begin{equation}
w({\mathbb H}^\gamma, {\mathbb H}^\delta) = w({\bf Q}{\mathbb H}^\gamma, {\bf Q}{\mathbb H}^\delta) \label{ied1}
\end{equation}
for arbitrary ${\bf Q} \in Orth^+$ (cf. Subsection \ref{ssm}). Tensors ${\mathbb H}^\alpha$ ($\alpha = \{\gamma,\delta\}$) admit polar decomposition ${\mathbb H}^\alpha = {\bf R}^\alpha {\mathbb U}^\alpha$, where ${\bf R}^\alpha \in Orth$ are non-unique and positive semidefinite tensors ${\mathbb U}^\alpha \in Sym$ are unique (cf. \eqref{sd8.18} and \eqref{sd8.19}); ${\mathbb U}^\alpha$ satisfy ${\mathbb U}^\alpha = {\mathbbm 1}^\alpha {\mathbb U}^\alpha {\mathbbm 1}^\alpha$. Define ${\mathbb R}^\alpha = {\mathbb H}^\alpha \left({\mathbb U}^\alpha\right)^{-1}$, where $\left({\mathbb U}^\alpha\right)^{-1}$ is the pseudoinverse of ${\mathbb U}^\alpha$ such that $\left({\mathbb U}^\alpha\right)^{-1}{\mathbb U}^\alpha = {\mathbb U}^\alpha \left({\mathbb U}^\alpha\right)^{-1} = \mathbbm{1}^\alpha$. Consequently relations ${\mathbb R}^\alpha = {\bf R}^\alpha {\mathbbm 1}^\alpha$ and $\left({\mathbb R}^\alpha\right)^{-1} = \left({\mathbb R}^\alpha\right)^T$ hold, where $\left({\mathbb R}^\alpha\right)^{-1}$ is the pseudoinverse of ${\mathbb R}^\alpha$ satisfying $\left({\mathbb R}^\alpha\right)^{-1} {\mathbb R}^\alpha = \mathbbm{1}^\alpha$ and ${\mathbb R}^\alpha \left({\mathbb R}^\alpha\right)^{-1} = \bar{\mathbbm{1}}$. Therefore, we also have ${\mathbb R}^\alpha \left({\mathbb R}^\alpha\right)^T = \bar{\mathbbm 1}$ and $\left({\mathbb R}^\alpha\right)^T {\mathbb R}^\alpha  = {\mathbbm 1}^\alpha$. To this end, note the decomposition of interface elastic distortion tensors
\begin{equation}
{\mathbb H}^\alpha = {\mathbb R}^\alpha {\mathbb U}^\alpha \label{ied2}
\end{equation}
into two unique tensors.

Choose ${\bf Q} = ({\mathbb R}^\gamma)^T + {\mathbb N}^\gamma \otimes {\mathbbm n}$; thus $\det {\bf Q} = \det {\bf R}^\gamma ({\bf R}^\gamma {\mathbb N}^\gamma \cdot {\mathbbm n}) = (\det {\bf R}^\gamma)^2 (= 1)$, where the first equality follows upon substituting ${\mathbb R}^\gamma = {\bf R}^\gamma - {\bf R}^\gamma {\mathbb N}^\gamma \otimes {\mathbb N}^\gamma$ in the expression for ${\bf Q}$ and then using identity $(1.1.6)$ from \cite{Silhavy97}. To prove the second equality, recall \eqref{epd13}$_1$ to write
\begin{equation}
{\mathbb R}^\gamma {\mathbb U}^\gamma {\mathbbm t}_1 \otimes {\mathbb R}^\gamma {\mathbb U}^\gamma {\mathbbm t}_2 = |({\mathbb H}^\gamma)^* {\mathbb N}^\gamma|  {\mathbbm n}
\end{equation}
and let ${\mathbbm t}_a \in T_\mathcal{N}^\gamma$ ($a = \{1,2\}$) be the two principal vectors of ${\mathbb U}^\gamma$ (the third one is given by ${\mathbb N}^\gamma$) such that $\{{\mathbbm t}_1,{\mathbbm t}_2,{\mathbb N}^\gamma \}$ forms a positively oriented orthogonal basis. The details are left to the reader.

Substitute the assumed ${\bf Q}$ in \eqref{ied1} to obtain
\begin{equation}
w = \hat{w}({\mathbb U}^\gamma, ({\mathbb R}^\gamma)^T{\mathbb H}^\delta) \label{ied3}
\end{equation}
or equivalently
\begin{equation}
w = \bar{w}({\mathbb C}^\gamma, {\mathbb M}), \label{ied5}
\end{equation}
where $\mathbb{C}^\gamma = (\mathbb{U}^{\gamma})^ 2$ and ${\mathbb M} = ({\mathbb H}^\gamma)^{-1} {\mathbb H}^\delta$ ($=({\mathbb K}^\gamma)^{-1} {\mathbb K}^\delta$) is the relative elastic (or plastic) distortion between two local configurations at a material point on the incoherent interface, cf. \eqref{ied4}. Tensor $\mathbb{M}$ is a linear map from $T_\mathcal{N}^\delta$ to $T_\mathcal{N}^\gamma$ and is related to true interface dislocation densities, as shown in \eqref{pwsl38.3}.

Additional insight is furnished by rewriting the dissipation inequality \eqref{tdi17} with the interfacial energy density given by \eqref{ied5}. Recall \eqref{tdi4} and define, cf. \eqref{tdi11},
\begin{equation}
j_s^\gamma {\mathbb T}^\gamma = w_{\mathbb{H}^\gamma} ({\mathbb{H}^\gamma})^T~\text{and}~j_s^\gamma {\mathbb T}^\delta= w_{\mathbb{H}^\delta} ({\mathbb{H}^\delta})^T   \label{ied8}
\end{equation}
such that ${\mathbb T} = {\mathbb T}^\gamma + {\mathbb T}^\delta$. The Cauchy stress tensor $\mathbb{T}$ is symmetric (cf. \eqref{bl26.4}$_2$) and satisfies $\mathbb{T} = \bar{\mathbbm{1}} \mathbb{T} \bar{\mathbbm{1}}$. Unlike $\mathbb{T}$, $\mathbb{T}^\alpha$ ($\alpha = \{\gamma,\delta\}$) are not symmetric. Their asymmetric parts are related as $Skw(\mathbb{T}^\gamma) = -Skw(\mathbb{T}^\delta)$ (since $Skw(\mathbb{T}) = {\bf 0}$), where $Skw()$ denotes the skew-symmetric part of the tensor (similarly let $Sym()$ denote the symmetric part of the tensor). Noting that $\bar{w}_{\mathbb{M}} = \mathbbm{1}^\gamma \bar{w}_{\mathbb{M}} \mathbbm{1}^\delta$ and $\bar{w}_{\mathbb{C}^\gamma} = \mathbbm{1}^\gamma \bar{w}_{\mathbb{C}^\gamma} \mathbbm{1}^\gamma$  (cf. \eqref{deronsur2}$_1$) we have, from \eqref{ied5} and \eqref{ied4}$_1$,
\begin{eqnarray}
\mathring{\bar{w}} &=& Sym(\bar{w}_{\mathbb{C}^\gamma}) \cdot \mathring{\mathbb{C}}^\gamma + \bar{w}_{\mathbb{M}} \cdot \mathring{\mathbb{M}} \nonumber
\\
&=& \left( 2 \mathbb{H}^\gamma Sym(\bar{w}_{\mathbb{C}^\gamma}) - (\mathbb{H}^\gamma)^{-T} \bar{w}_{\mathbb{M}} \mathbb{M}^T \right)  \cdot \mathring{\mathbb{H}}^\gamma + (\mathbb{H}^\gamma)^{-T} \bar{w}_{\mathbb{M}} \cdot \mathring{\mathbb{H}}^\delta. \label{ied5.1}
\end{eqnarray}
The coefficient of $\mathring{\mathbb{H}}^\alpha$ above should be equal to  $w_{\mathbb{H}^\alpha}$ for each $\alpha = \{\gamma,\delta\}$. Exploiting this correspondence and using \eqref{ied8} we get
\begin{equation}
j_s^\gamma \mathbb{T} = 2 \mathbb{H}^\gamma Sym(\bar{w}_{\mathbb{C}^\gamma}) (\mathbb{H}^\gamma)^{T}~\text{and}~ j_s^\gamma {\mathbb T}^\delta= (\mathbb{H}^\gamma)^{-T} \bar{w}_{\mathbb{M}} ({\mathbb{H}^\delta})^T. \label{ied5.2}
\end{equation}
Therefore, the dependence of $\bar{w}$ on $\mathbb{C}^\gamma$ alone contributes to the total interface stress; and the dependence on $\mathbb{M}$ to the dissipation, as shown below. Use \eqref{bl26.3}, \eqref{tdi6}, and \eqref{tdi11}$_2$ to obtain
\begin{equation}
j^\gamma \mathbb{P} = 2 \mathbb{H}^\gamma Sym(\bar{w}_{\mathbb{C}^\gamma}) (\mathbb{K}^\gamma)^{T}~\text{and}~ j^\gamma {\mathbb P}^\delta= (\mathbb{H}^\gamma)^{-T} \bar{w}_{\mathbb{M}} ({\mathbb{K}^\delta})^T \label{ied5.3}
\end{equation}
for interface Piola stresses. On the other hand, use the normal time derivative of  \eqref{ied4}$_2$ and the definition \eqref{tdi15} for $\mathbb{E}^\delta$ to show
\begin{equation}
\mathbb{E}^\delta \cdot \mathring{\mathbb{K}}^\delta ({\mathbb{K}^\delta})^{-1} = \mathbb{E}^\delta \cdot \left( \mathbb{K}^\gamma \mathring{\mathbb{M}} ({\mathbb{K}^\delta})^{-1} + \mathring{\mathbb{K}}^\gamma ({\mathbb{K}^\gamma})^{-1} \right) \label{ied5.4}
\end{equation}
and thus
\begin{equation}
\sum_{\alpha=\gamma,\delta} \mathbb{E}^\alpha \cdot \mathring{\mathbb{K}}^\alpha({\mathbb{K}^\alpha})^{-1} = \mathbb{E} \cdot \mathring{\mathbb{K}}^\gamma ({\mathbb{K}^\gamma})^{-1} - (j^\gamma)^{-1} \bar{w}_{\mathbb{M}} \cdot \mathring{\mathbb{M}}, \label{ied5.5}
\end{equation}
where we have used \eqref{tdi15}, \eqref{ied5.3}$_2$, and \eqref{epd8}. Substitute \eqref{ied5.5} into the dissipation inequality \eqref{tdi17} to obtain its alternate form
\begin{equation}
U \mathbbm{f} +  \mathbb{E} \cdot \mathring{\mathbb{K}}^\gamma ({\mathbb{K}^\gamma})^{-1} - (j^\gamma)^{-1} \bar{w}_{\mathbb{M}} \cdot \mathring{\mathbb{M}} \geq 0 ~  \forall {\bf
X} \in S_r, \label{tdi17a}
\end{equation}
where $\mathbbm{f}$ and $\mathbb{E}$ are given in \eqref{tdi18} and \eqref{cdi23.1}, respectively. Hence there are three dissipative processes active at an incoherent interface: the normal motion of the interface (driven by $\mathbbm{f}$), the evolution of plastic distortion (driven by $\mathbb{E}$), and the evolution of relative plastic (or elastic) distortion (driven by $(j^\gamma)^{-1} \bar{w}_{\mathbb{M}}$). The normal projections $\mathring{\mathbb{K}}^\gamma \mathbb{N}^\gamma$ and $\mathring{\mathbb{M}} \mathbb{N}^\delta$ do not contribute to the dissipation. If, however, the energy density depends explicitly on interfacial normals then these normal projections would also contribute to the dissipation, as discussed in a remark above. The dissipation inequality as given in \eqref{tdi17} or \eqref{tdi17a} has to be satisfied by all kinetic laws governing the irreversible process of coupled plastic flow for an interface moving within a solid.

Next, we note the restrictions imposed by material symmetry on the form of interface energy density. Recall the discussion in Subsection \ref{ssm} and consider ${\bf G}^\gamma \in \mathcal{G}^\gamma$ and ${\bf G}^\delta \in \mathcal{G}^\delta$, where $\mathcal{G}^\gamma$ and $\mathcal{G}^\delta$ are the symmetry groups associated with $T_\mathcal{N}^\gamma$ and $T_\mathcal{N}^\delta$, respectively. The material response at the interface remains invariant under the action of these groups. Therefore
\begin{equation}
w({\mathbb H}^\gamma, {\mathbb H}^\delta) = w({\mathbb H}^\gamma {\mathbb G}^\gamma, {\mathbb H}^\delta {\mathbb G}^\delta), \label{ied6}
\end{equation}
where $\mathbb{G}^\alpha = \mathbbm{1}^\alpha {\bf G}^\alpha$, cf. \eqref{invcomp1.031} and \eqref{invcomp1.05}. For the energy density given by \eqref{ied5} we require
\begin{equation}
\bar{w}({\mathbb C}^\gamma, {\mathbb M}) = \bar{w}(({\mathbb G}^\gamma)^T {\mathbb C}^\gamma {\mathbb G}^\gamma, ({\mathbb G}^\gamma)^T{\mathbb M} {\mathbb G}^\delta) \label{ied7}
\end{equation}
to be satisfied by all elements of symmetry groups $\mathcal{G}^\gamma$ and $\mathcal{G}^\delta$. Indeed, under the action of these symmetry groups ${\mathbb C}^\gamma = ({\mathbb H}^\gamma)^T {\mathbb H}^\gamma$ transforms to $({\mathbb G}^\gamma)^T {\mathbb C}^\gamma {\mathbb G}^\gamma$, cf. \eqref{invcomp1.031}; and ${\mathbb M} = ({\mathbb H}^\gamma)^{-1} {\mathbb H}^\delta$ transforms to $({\mathbb G}^\gamma)^{-1}{\mathbb M}{\mathbb G}^\delta$, where $({\mathbb G}^\gamma)^{-1}$ is the pseudoinverse of ${\mathbb G}^\gamma$ satisfying $\left({\mathbb G}^\gamma\right)^{-1} {\mathbb G}^\gamma = \mathbbm{1}^{\gamma'}$ and ${\mathbb G}^\gamma \left({\mathbb G}^\gamma\right)^{-1} = \mathbbm{1}^\gamma$ (here $\mathbbm{1}^{\gamma'}$ represents the surface identity tensor for the transformed local configuration at the $\gamma$-surface). Finally, recall \eqref{invcomp1.05} and the uniqueness of the pseudoinverse to note that $\left({\mathbb G}^\gamma\right)^{-1} = \left({\mathbb G}^\gamma\right)^{T}$. These considerations provide a point of departure for the phenomenological modeling of the behavior of incoherent interfaces which, contrary to coherent boundaries, are affected by two possibly distinct symmetry groups.

\begin{rem} The skew components of $\mathbb{T}^\alpha$ are conjugate to the relative spin tensor across the interface. To elaborate, we write the normal time derivative of $w$ in the form (using \eqref{ied8})
\begin{equation}
j_s^\gamma \mathring{w} = \sum_{\alpha=\gamma,\delta}  \mathbb{T}^\alpha \cdot  \mathrm{L}^\alpha, \label{ied10}
\end{equation}
where $\mathrm{L}^\alpha = \mathring{\mathbb{H}}^\alpha ({\mathbb{H}^\alpha})^{-1}$. Decompose $\mathrm{L}^\alpha$ into symmetric ($\mathrm{D}^\alpha$) and skew ($\mathrm{W}^\alpha$) parts. The tensors $\mathrm{D}^\alpha$ and $\mathrm{W}^\alpha$ are identified as the stretching tensor and the spin tensor, respectively. Substitute the decomposition in \eqref{ied10} to get
\begin{eqnarray}
j_s^\gamma \mathring{w} &=& \sum_{\alpha=\gamma,\delta} \Big( Sym(\mathbb{T}^\alpha) \cdot  \mathrm{D}^\alpha + Skw(\mathbb{T}^\alpha) \cdot  \mathrm{W}^\alpha \Big) \nonumber
\\
&=& \sum_{\alpha=\gamma,\delta} Sym(\mathbb{T}^\alpha) \cdot  \mathrm{D}^\alpha + Skw(\mathbb{T}^\delta) \cdot  \left(\mathrm{W}^\delta - \mathrm{W}^\gamma \right), \label{ied11}
\end{eqnarray}
where the second equality is a result of the symmetry of $\mathbb{T}$. Finally, note that $Skw(\mathbb{T}^\gamma)$ is completely determined from $Skw(\bar{w}_{\mathbb{M}})$, cf. \eqref{ied5.2}$_2$.
\end{rem}

\begin{rem}
If the interface is unstrained, i.e. $\mathbb{H}^\alpha = \mathbb{R}^\alpha$ (for $\{\alpha=\gamma,\delta\}$), then the representation \eqref{ied5}
reduces to
\begin{equation}
w = \bar{w}({\mathbbm 1}^\gamma, {\mathbb R}), \label{sed5}
\end{equation}
where $\mathbb{R} = (\mathbb{R}^\gamma)^T \mathbb{R}^\delta$ is the relative rotation at the interface. Such energies
are widely studied in the literature on metal interfaces, in
particular grain boundaries, and many experimental methods have been
devised for their evaluation.
\end{rem}

\subsection{\label{kr} Kinetic relations}

Motivated by the dissipation inequality \eqref{tdi17} we consider constitutive functions of the form
\begin{equation}
f = \hat{f}({\mathbbm f}, {\mathbb H}^\alpha, {\mathbb K}^\alpha, \mathring{\mathbb H}^\alpha, \mathring{\mathbb K}^\alpha, \mathbb{L}, \mathbb{N}^\alpha, U), \label{inv18.9}
\end{equation}
where the arguments of $\hat{f}$ can include both $\gamma$ and $\delta$ variables ($\alpha = \{\gamma,\delta\}$). Relations of the type $f=0$ would then furnish possible candidates for kinetic laws at the interface. These relations complete the set of equations necessary to determine the evolution of state variables.

We first obtain restrictions on $f$ to be invariant under compatible changes in the reference configuration. Recall that a compatible change ensures that the Euclidean nature of the reference configuration remains unaltered. Consider, as in Subsections \ref{ssm} and \ref{trsurdd}, two reference configurations $\kappa_{r_1}$ and $\kappa_{r_2}$ related by a map ${\mbf \lambda}$ such that ${\bf X}_2 = {\mbf \lambda}({\bf X}_1)$, where ${\bf X}_1 \in \kappa_{r_1}$ and ${\bf X}_2 \in \kappa_{r_2}$, with continuous ${\bf A} \in InvLin$  given by ${\bf A} = \nabla_1 {\mbf \lambda}$. As before, we assume that the transformation leaves the normal time derivative invariant and that $\mathring{\mathbb A} = {\bf 0}$. They imply conditions \eqref{invcomp1.011} and \eqref{invcomp1.012}, respectively.

The function $f$ is invariant under a change of reference configuration from $\kappa_{r_1}$ to $\kappa_{r_2}$ if
\begin{equation}
\hat{f}({\mathbbm f}_1, {\mathbb H}_1^\alpha, {\mathbb K}_1^\alpha, \mathring{\mathbb H}^\alpha_1, \mathring{\mathbb K}^\alpha_1, \mathbb{L}_1, \mathbb{N}^\alpha_1, U_1) = \hat{f}({\mathbbm f}_2, {\mathbb H}_2^\alpha, {\mathbb K}_2^\alpha, \mathring{\mathbb H}^\alpha_2, \mathring{\mathbb K}^\alpha_2, \mathbb{L}_2, \mathbb{N}^\alpha_2, U_2), \label{inv19}
\end{equation}
where all normal time derivatives are given with respect to $\kappa_{r_1}$.

The transformations for bulk and interface distortions are given in \eqref{invcomp1} and \eqref{invcomp1.02}; and for the normal velocity, normal, and curvature tensor in \eqref{inv16}, \eqref{invcomp1.013}, and \eqref{inv18.1}, respectively. Substitute ${\bf A} = {\bf K}_2^\pm ({\bf K}_1^\pm)^{-1}$ and $j_A = {j_2^\alpha}({j_1^\alpha})^{-1}$ (no summation over $\alpha = \{\gamma, \beta\}$) into \eqref{inv16} and \eqref{inv18.1} to get $j_2^\alpha J_{K_2^\pm}^{-1}  U_2 = j_1^\alpha J_{K_1^\pm}^{-1} U_1$ and ${j_2^\alpha}^{-1} J_{K_2^\pm} ({\mathbb K}_2^\alpha)^{-1} \mathbb{L}_2 {\mathbb K}_2^\alpha= {j_1^\alpha}^{-1} J_{K_1^\pm} ({\mathbb K}_1^\alpha)^{-1} \mathbb{L}_1 {\mathbb K}_1^\alpha$; thus define
\begin{equation}
{U}^\alpha = h^\alpha U~\text{and}~\mathbb{L}^\alpha = (h^\alpha)^{-1} ({\mathbb K}^\alpha)^{-1} \mathbb{L} {\mathbb K}^\alpha, \label{inv18}
\end{equation}
where $h^\alpha = j^\alpha J_{K^\pm}^{-1}$ (superscript $+$ is used for defining  $h^\gamma$  and $-$ for $h^\delta$), as invariant (or true) normal speeds and invariant (or true) curvature tensors associated with the singular interface.

The transformation in the driving force $\mathbbm{f}$ can be evaluated by recalling its expression from \eqref{tdi18} and noting that
\begin{eqnarray}
{\bf E}_2 &=& J_A^{-1} {\bf A}^{-T} {\bf E}_1 {\bf A}^T, ~\rho_{r_2} = J_A^{-1} \rho_{r_1}, ~\text{and} \label{inv19.1}
\\
{\mathbb E}_2 &=& j_A^{-1} {\mathbb A}^{-T} {\mathbb E}_1 {\mathbb A}^T. \label{inv19.2}
\end{eqnarray}
These can be proved using \eqref{cdi23}, \eqref{cdi23.1}, \eqref{tdi5}, \eqref{tdi6}, and \eqref{invcomp1}$-$\eqref{invcomp1.03} with the fact that $\Psi$ and $\rho_r$ are densities per unit volume of $\kappa_r$ and $\Phi$ is a density per unit area of $S_{r}$. Combine them with \eqref{inv16}, \eqref{invcomp1.013}, and \eqref{inv18.1} to get
\begin{equation}
{\mathbbm f}_2 = J_A^{-1} {\mathbbm f}_1 \label{inv19.3}
\end{equation}
and hence define $\mathbbm{f}^\alpha = J_{K^\pm} {\mathbbm f}$ as the invariant driving force for normal motion of the interface (superscript $+$ is used for defining  $\mathbbm{f}^\gamma$  and $-$ for $\mathbbm{f}^\delta$).

To obtain a necessary condition for \eqref{inv19}, let ${\bf A} = ({\bf K}_1^+)^{-1}$ (therefore $\mathbb{A} = (\mathbb{K}_1^\gamma)^{-1}$) locally at the point at which \eqref{inv19} is evaluated. With this choice for ${\bf A}$, relations \eqref{invcomp1}, \eqref{invcomp1.02}, \eqref{inv19.3}, \eqref{inv16}, and \eqref{inv18.1} yield
\begin{eqnarray}
&& {\mathbb K}_2^\gamma = {\mathbbm 1}^\gamma,~ {\mathbb K}_2^\delta = {\mathbb M}_1, \label{inv20}
\\
&& \mathring{\mathbb K}^\gamma_2 = (\mathbb{K}_1^\gamma)^{-1} \mathring{\mathbb K}^\gamma_1,~  \mathring{\mathbb K}^\delta_2 = (\mathbb{K}_1^\gamma)^{-1} \mathring{\mathbb K}^\delta_1 \label{inv21}
\\
&& {\mathbbm f}_2 = J_{K_1^+} {\mathbbm f}_1,~ U_2 =  h_1^\gamma U_1,~ \text{and} \label{inv22}
\\
&& \mathbb{L}_2 = (h_1^\gamma)^{-1} ({\mathbb K}_1^\gamma)^{-1} \mathbb{L}_1 {\mathbb K}_1^\gamma,   \label{inv22.1}
\end{eqnarray}
whereas ${\mathbb H}^\alpha$, $\mathring{\mathbb H}^\alpha$, and $\mathbb{N}^\alpha$ remain invariant. The above argument can be repeated with $-$ superscript in place of  $+$ (and $\delta$ in place of $\gamma$). We therefore obtain the necessary and sufficient condition for $f$ to be invariant under compatible changes in the reference configuration as
\begin{equation}
f = \breve{f}(\mathbbm{f}^\alpha, {\mathbb H}^\alpha, \mathbb{M}, \mathring{\mathbb H}^\alpha, (\mathbb{K}^\gamma)^{-1} \mathring{\mathbb K}^\alpha, \mathbb{L}^\alpha, \mathbb{N}^\alpha, U^\alpha). \label{inv27}
\end{equation}
The proof of sufficiency is straightforward and therefore omitted.

The list of arguments can be reduced on noting the definition \eqref{ied4} of $\mathbb{M}$ in addition to
\begin{eqnarray}
&& {\mathbb N}^\gamma = m^{-1} \mathbb{M}^* {\mathbb N}^\delta,~\mathbbm{f}^\delta = \frac{J_{K^-}}{J_{K^+}} \mathbbm{f}^\gamma,~U^\gamma = \frac{J_{K^-}}{J_{K^+}} m^{-1} U^\delta,  \label{inv28}
\\
&& (\mathbb{K}^\gamma)^{-1} \mathring{\mathbb K}^\delta = \mathbbm{1}^\gamma \mathring{\mathbb M} + (\mathbb{K}^\gamma)^{-1} \mathring{\mathbb K}^\gamma \mathbb{M}, ~\text{and} \label{inv29}
\\
&& {\mathbb L}^\delta = \frac{J_{K^-}}{J_{K^+}} m^{-1} \mathbb{M}^{-1} \mathbb{L}^\gamma \mathbb{M}, \label{inv30}
\end{eqnarray}
where $m = |\mathbb{M}^*| = j^\delta (j^\gamma)^{-1}$.
The representation \eqref{inv27} can then be equivalently written in the form
\begin{equation}
f = \tilde{f}(\mathbbm{f}^\alpha, {\mathbb H}^\gamma, \mathbb{M}, \mathring{\mathbb H}^\alpha, \mathring{\mathbb M}, (\mathbb{K}^\gamma)^{-1} \mathring{\mathbb K}^\gamma, \mathbb{L}^\gamma, \mathbb{N}^\gamma, U^\gamma). \label{inv32}
\end{equation}
This provides us with a complete and mutually independent set of variables that can be used as arguments for constitutive functions which are invariant under compatible transformation of the reference configuration.

Under superimposed rigid body motions, all except ${\mathbb H}^\gamma$ and $\mathring{\mathbb H}^\gamma$ in the arguments of $\tilde{f}$ remain invariant. They transform to ${\bf Q} {\mathbb H}^\gamma$ and ${\bf Q}\mathring{\mathbb H}^\gamma + \mathring{\bf Q} {\mathbb H}^\gamma$ respectively, where ${\bf Q} \in Orth^+$ is arbitrary, cf. Subsection \ref{ssm}. For $\tilde{f}$ to be invariant under superimposed rigid body motions we require
\begin{equation}
\tilde{f}(\mathbb{H}^\gamma, \mathring{\mathbb H}^\gamma, \cdots) = \tilde{f}({\bf Q} {\mathbb H}^\gamma, {\bf Q}\mathring{\mathbb H}^\gamma + \mathring{\bf Q} {\mathbb H}^\gamma, \cdots), \label{inv34}
\end{equation}
where dependence on other variables is suppressed. Choose ${\bf Q} = ({\mathbb R}^\gamma)^T + {\mathbb N}^\gamma \otimes {\mathbbm n}$ where ${\mathbb R}^\gamma$ is defined in \eqref{ied2}. With this choice ${\bf Q} {\mathbb H}^\gamma = \mathbb{U}^\gamma$ and ${\bf Q}\mathring{\mathbb H}^\gamma + \mathring{\bf Q} {\mathbb H}^\gamma = \mathring{\mathbb{U}}^\gamma$. The second of these follows on using $(\mathbb{R}^\gamma)^T \mathbb{R}^\gamma = {\mathbbm 1}^\gamma$, $(\mathbb{R}^\gamma)^T {\mathbbm n} = {\bf 0}$, $\mathbb{U}^\gamma {\mathbb N}^\gamma = {\bf 0}$, and their normal time derivatives. Recalling $\mathbb{C}^\gamma = ({\mathbb U}^\gamma)^2$ we thus have the necessary and sufficient condition for invariance under rigid body motions:
\begin{equation}
f = \bar{f}(\mathbbm{f}^\alpha, {\mathbb C}^\gamma, \mathbb{M}, \mathring{\mathbb C}^\gamma, \mathring{\mathbb M}, (\mathbb{K}^\gamma)^{-1} \mathring{\mathbb K}^\gamma, \mathbb{L}^\gamma, \mathbb{N}^\gamma, U^\gamma). \label{inv35}
\end{equation}
The proof for sufficiency is straightforward and therefore omitted.

Assume for illustrative purposes that the conditions required by inverse function theorem are satisfied such that we can invert $f = 0$, where $f$ is as given in \eqref{inv35}, to obtain
\begin{eqnarray}
&& U^\gamma = \bar{U}^\gamma (\mathbbm{f}^\alpha, {\mathbb C}^\gamma, \mathbb{M}, \mathring{\mathbb C}^\gamma, \mathring{\mathbb M}, (\mathbb{K}^\gamma)^{-1} \mathring{\mathbb K}^\gamma, \mathbb{L}^\gamma, \mathbb{N}^\gamma), \label{inv35.1}
\\
&& \mathring{\mathbb{M}} = \mathcal{R}(\mathbbm{f}^\alpha, {\mathbb C}^\gamma, \mathbb{M}, \mathring{\mathbb C}^\gamma, (\mathbb{K}^\gamma)^{-1} \mathring{\mathbb K}^\gamma, \mathbb{L}^\gamma, \mathbb{N}^\gamma, U^\gamma), ~\text{and} \label{inv36}
\\
&& (\mathbb{K}^\gamma)^{-1} \mathring{\mathbb K}^\gamma = \mathcal{S}(\mathbbm{f}^\alpha, {\mathbb C}^\gamma, \mathbb{M}, \mathring{\mathbb C}^\gamma, \mathring{\mathbb M}, \mathbb{L}^\gamma, \mathbb{N}^\gamma, U^\gamma). \label{inv37}
\end{eqnarray}
The following restrictions imposed by the material symmetry on $\bar{U}^\gamma$, $\mathcal{R}$, and $\mathcal{S}$ ensue on recalling the pertinent discussion from Subsection \ref{ssm}:
\begin{eqnarray}
&& \bar{U}^\gamma  = \bar{U}^\gamma(\mathbbm{f}^\alpha, \hat{\mathbb C}^\gamma, \hat{\mathbb{M}}, \mathring{\hat{\mathbb C}}^\gamma, \mathring{\hat{\mathbb M}}, (\hat{\mathbb{K}}^\gamma)^{-1} \mathring{\hat{\mathbb K}}^\gamma, \hat{\mathbb{L}}^\gamma, \hat{\mathbb{N}}^\gamma ),  \label{inv37.1}
\\
&& ({\mathbb G}^\gamma)^T \mathcal{R} {\mathbb G}^\delta  = \mathcal{R}( \mathbbm{f}^\alpha, \hat{\mathbb C}^\gamma, \hat{\mathbb{M}}, \mathring{\hat{\mathbb C}}^\gamma, (\hat{\mathbb{K}}^\gamma)^{-1} \mathring{\hat{\mathbb K}}^\gamma, \hat{\mathbb{L}}^\gamma, \hat{\mathbb{N}}^\gamma, U^\gamma ),  \label{inv38}
\\
&& \text{and}~~({\mathbb G}^\gamma)^T \mathcal{S} {\mathbb G}^\gamma = \mathcal{S}( \mathbbm{f}^\alpha, \hat{\mathbb C}^\gamma, \hat{\mathbb{M}}, \mathring{\hat{\mathbb C}}^\gamma, \mathring{\hat{\mathbb M}}, \hat{\mathbb{L}}^\gamma, \hat{\mathbb{N}}^\gamma, U^\gamma ), \label{inv39}
\end{eqnarray}
where (recall $\mathbb{G}^\alpha = \mathbbm{1}^\alpha {\bf G}^\alpha$, \eqref{invcomp1.031}, and \eqref{invcomp1.05})
\begin{eqnarray}
&& \hat{\mathbb C}^\gamma = ({\mathbb G}^\gamma)^T\mathbb{C}^\gamma {\mathbb G}^\gamma,~\hat{\mathbb{M}} = ({\mathbb G}^\gamma)^T \mathbb{M}{\mathbb G}^\delta, \label{inv40}
\\
&& \mathring{\hat{\mathbb C}}^\gamma = ({\mathbb G}^\gamma)^T \mathring{\mathbb C}^\gamma {\mathbb G}^\gamma,~\mathring{\hat{\mathbb M}} = ({\mathbb G}^\gamma)^T\mathring{\mathbb{M}} {\mathbb G}^\delta,~(\hat{\mathbb{K}}^\gamma)^{-1} \mathring{\hat{\mathbb K}}^\gamma = ({\mathbb G}^\gamma)^T(\mathbb{K}^\gamma)^{-1} \mathring{\mathbb K}^\gamma {\mathbb G}^\gamma, \label{inv41}
\\
&& \hat{\mathbb{L}}^\gamma = ({\mathbb G}^\gamma)^T \mathbb{L}^\gamma {\mathbb G}^\gamma,~\text{and}~ \hat{\mathbb{N}}^\gamma = ({\bf G}^\gamma)^T \mathbb{N}^\gamma. \label{inv42}
\end{eqnarray}
Here ${\bf G}^\gamma \in \mathcal{G}^\gamma$ and ${\bf G}^\delta \in \mathcal{G}^\delta$ are the symmetry maps at the interface. The scalars $\mathbbm{f}^\alpha$ and $U^\alpha$ remain invariant since $J_{G^\alpha} = 1$ and $| ({\mathbb G}^\alpha)^* | = 1$.

\begin{rem} If $\bar{f}$ is independent of $\mathring{\mathbb{N}}^\gamma$ and $\mathring{\mathbb{N}}^\delta$ then its dependence on $\mathring{\mathbb C}^\gamma$, $\mathring{\mathbb{M}}$, and $\mathring{\mathbb K}^\gamma$ is only through ${\mathbbm{1}}^\gamma  \mathring{\mathbb C}^\gamma {\mathbbm{1}}^\gamma $, ${\mathbbm{1}}^\gamma \mathring{\mathbb{M}} {\mathbbm{1}}^\delta$, and $\mathring{\mathbb K}^\gamma {\mathbbm{1}}^\gamma$, respectively. These follow immediately on taking the normal time derivative of  ${\mathbb C}^\gamma \mathbb{N}^\gamma = {\bf 0}$, ${\mathbb{M}}^T\mathbb{N}^\gamma = {\bf 0}$, ${\mathbb{M}} \mathbb{N}^\delta = {\bf 0}$, and ${\mathbb K}^\gamma \mathbb{N}^\gamma = {\bf 0}$.
\end{rem}

\begin{rem} A boundary-initial-value problem for ${\mbf \chi}({\bf X}, t)$ and ${\bf K}({\bf X}, t)$ is specified through the coupled system of nonlinear partial differential equations given by \eqref{bl18} (with $\bf P$ substituted from \eqref{tdi5}) and a flow rule of the type (cf. equation $(102)$ of \cite{Guptaetal07})
\begin{equation}
\dot{{\bf K}^{-1}} {\bf K}=\mathcal{H}({\bf C}_{H}, \dot{\bf C}_{H}, {\mbf \alpha}), \label{fay6}
\end{equation}
where ${\bf C}_H = {\bf H}^T {\bf H}$ and ${\mbf \alpha}$ is the bulk dislocation density defined in \eqref{dai6}. The boundary data on $S_r$ are furnished by \eqref{bl19} combined with the interfacial kinetic laws \eqref{inv35.1}$-$\eqref{inv37}. The problem is completed by supplying appropriate boundary conditions on $\partial \kappa_r$ and initial conditions for ${\mbf \chi}$ and $\bf K$ at some fixed time.
\end{rem}

{\small Anurag Gupta acknowledges the support of the initiation grant from Indian Institute of Technology, Kanpur, India.}

\bibliography{Thesis_AnuragPlas}
\bibliographystyle{plain}

\end{document}